\documentclass[aps,prd,amssymb,amsmath,nobibnotes,nofootinbib,superscriptaddress, preprint,12pt]{revtex4}
\usepackage{amsfonts,amsmath,hyperref,url, color}
\allowdisplaybreaks
\usepackage{bm, bbm}
\usepackage{graphicx}
\usepackage{mathtools}
\usepackage{t1enc}
\usepackage{hepnames}
\usepackage{tensor}
\usepackage{comment}

\newcommand{\bc}{\begin{center}}
\newcommand{\ec}{\end{center}}
\newcommand{\be}{\begin{eqnarray}}
\newcommand{\ee}{\end{eqnarray}}
\newcommand{\bs}{\begin{slide}}
\newcommand{\es}{\end{slide}}

\newcommand{\bi}{\begin{itemize}}
\newcommand{\ei}{\end{itemize}}
\newcommand{\nn}{\nonumber}

\begin{document}
\title{$\kappa$-deformed complex fields and discrete symmetries}

\author{Michele Arzano}
\affiliation{Dipartimento di Fisica ``E. Pancini" and INFN, Universit\`a di Napoli Federico II, Via Cinthia,
80126 Fuorigrotta, Napoli, Italy}
\author{Andrea Bevilacqua}
\affiliation{National Centre for Nuclear Research, ul. Pasteura 7, 02-093 Warsaw, Poland}
\author{Jerzy Kowalski-Glikman}
\affiliation{Institute for Theoretical Physics, University of Wroc\l{}aw, pl.\ M.\ Borna 9, 50-204
Wroc\l{}aw, Poland}
\affiliation{National Centre for Nuclear Research, ul. Pasteura 7, 02-093 Warsaw, Poland}
\author{Giacomo Rosati}
\affiliation{Institute for Theoretical Physics, University of Wroc\l{}aw, pl.\ M.\ Borna 9, 50-204
Wroc\l{}aw, Poland}
\author{Josua Unger}
\affiliation{Institute for Theoretical Physics, University of Wroc\l{}aw, pl.\ M.\ Borna 9, 50-204
Wroc\l{}aw, Poland}

\date{\today}

\begin{abstract}
We present a construction of $\kappa$-deformed complex scalar field theory with the objective of shedding light on the way discrete symmetries and CPT invariance are affected by the deformation. Our starting point is the observation that, in order to have an appropriate action of Lorentz symmetries on antiparticle states, these should be described by four-momenta living on the complement of the portion of de Sitter group manifold to which $\kappa$-deformed particle four-momenta belong. Once the equations of motions are properly worked out from the deformed action we obtain that particle and antiparticle are characterized by different mass-shell constraints leading to a subtle form of departure from CPT invariance.The remaining part of our work is dedicated to a detailed description of the action of deformed Poincar\'e and discrete symmetries on the complex field.

\end{abstract}`

\maketitle

\section{Introduction}

It is commonly expected that the usual description of space-time as a smooth manifold is no longer reliable as we approach the Planck scale when quantum effects of the geometry can no longer be neglected. Since the pre-history of research on quantum gravity\footnote{According to Jackiw \cite{Jackiw:2003dw} the idea of non-commuting space-time coordinates was first suggested by Heisenberg back in the 1930s. He then discussed it with Peierls who in turn told to Pauli who told Oppenheimer who asked his student Snyder to work it out in detail and thus the fist paper on non-commutative space-time was published in 1947 \cite{Snyder:1946qz}.} non-commutativity of space-time has been advocated  as a possible way to effectively model quantum gravitational effects in regimes of negligible curvature. A widely studied incarnation of this idea suggests that the scale of non-commutativity should be seen as an observer independent length scale \cite{AmelinoCamelia:2000mn}, and that, in order to accommodate such fundamental scale, ordinary relativistic symmetries should be {\it deformed} into non-trivial Hopf algebras which, in the limit of vanishing non-commutativity, should reproduce the usual Poincar\'e algebra.

The $\kappa$-Poincar\'e algebra is an example of such deformations which has been intensively investigated for almost 30 years. Such algebra was originally derived by contracting the quantum anti-de Sitter algebra  \cite{Lukierski:1991pn, Lukierski:1992dt}. It was brought to its modern form a few years later in \cite{Lukierski:1993wx} and \cite{Majid:1994cy}, where, in particular, the role of non-commutative $\kappa$-Minkowski spacetime was discovered and investigated. The deformation parameter $\kappa$ has dimensions of mass and, in light of the possible role of the $\kappa$-Poincar\'e algebra in describing the symmetries of a flat-space-time limit of quantum gravity, it is usually identified with the Planck energy. Such putative relationship with a semiclassical limit of quantum gravity renders this model especially relevant for the search of possible experimental signatures of Planck scale physics \cite{AmelinoCamelia:1999zc, AmelinoCamelia:2008qg}. So far most of the proposed observational frameworks having sufficient sensitivity to capture effects of quantum gravity origin \cite{AmelinoCamelia:2003ex, AmelinoCamelia:2008qg} were based on purely kinematical models, like, for example, the well known case of measuring the time of flight of Gamma Ray Bursts photons of different energies \cite{AmelinoCamelia:1997gz, Amelino-Camelia:2016ohi}. It has however been argued that $\kappa$-deformations may have a subtle, and in principle measurable, effect on elementary particles, linked to the deformation of CPT symmetry \cite{Arzano:2019toz}. For these reasons we believe that developing a comprehensive theory of deformed quantum fields will be beneficial for better understanding known phenomena related to $\kappa$-deformation and possibly shed light on some new ones that might be of phenomenological relevance (besides, of course, its relevance at a purely theoretical level).


In the series of papers, of which the present one is the first, we will formulate the theory of a free, complex $\kappa$-deformed scalar field. The next paper in the series will be devoted to free scalar field propagator and n-point functions. We will consider next massive higher-spin fields and then the  quantum deformed abelian gauge fields. We will discuss interacting fields in the final, fifth paper of the series.

The present paper has its roots in the work \cite{Freidel:2007hk} from which we borrow the notation and most of conventions. However there are important differences. In particular, the definition of the scalar field is different here. This change of definition is a consequence of the assumed nice behavior of the field with respect to the discrete CPT transformations and leads to one of the major results of this paper, that the mass shell relations of particles and antiparticles differ from each other, although as a manifold the mass-shell in both case is the same hyperboloid in momentum space, as anticipated in \cite{Arzano:2019toz}, \cite{Arzano:2020rzu}. Thanks to this new definition of fields also the creation-annihilation operator algebra becomes particularly simple. In the present paper we also consistently use the star product formalism instead of the equivalent formalism of non-commutative spacetime used in \cite{Freidel:2007hk}.

Various aspects of the theory of $\kappa$-deformed fields were discussed in the past. Here we mention papers that influenced us \cite{AmelinoCamelia:2001fd}--\cite{Mercati:2018hlc} in working on this project, but we would like to stress that the crucial aspects of the present construction, like the doubling of momentum space and insistence on the proper action of discrete symmetries are new.

\section{Preliminaries}
\label{sec:preliminaries}

As it is well known there are two complementary pictures of $\kappa$-deformation. One deals with the presence of non-commutative spacetime with Lie type non-commutativity, called $\kappa$-Minkowski space \cite{Lukierski:1993wx}, \cite{Majid:1994cy}, where the commutator of coordinates $\hat x^\mu$ form the $\mathfrak{an}(3)$ Lie  algebra
\begin{equation}\label{0.1}
 [\hat x^0, \hat x^i] = \frac{i}{\kappa}\,  \hat x^i\,,
\end{equation}
with the parameter $\kappa$ defining the `strength' of non-commutativity.
Another concerns the momentum space picture, in which the momentum space is curved and is a submanifold of de Sitter space with curvature $1/\kappa^2$ \cite{KowalskiGlikman:2002ft}, \cite{KowalskiGlikman:2003we}, which is constructed as follows.

Let us consider the following 5-dimensional
matrix representation of the  Lie algebra \eqref{0.1}
\begin{equation}\label{II.1.18}
\hat x^0 = -\frac{i}{\kappa} \,\left(\begin{array}{ccc}
  0 & \mathbf{0} & 1 \\
  \mathbf{0} & \mathbf{0} & \mathbf{0} \\
  1 & \mathbf{0} & 0
\end{array}\right) \quad
\hat{\mathbf{x}} = -\frac{i}{\kappa} \,\left(\begin{array}{ccc}
  0 & {\bm{\epsilon}\,{}^T} &  0\\
  \bm{\epsilon} & \mathbf{0} & \bm{\epsilon} \\
  0 & -\bm{\epsilon}\,{}^T & 0
\end{array}\right),
\end{equation}
where bold fonts are used to denote space components of a 4-vector (with the exception of the central $\bm{0}$ which is a $3\times 3$ matrix)
and $\bm{\epsilon}$ is a three dimensional vector with a single unit
entry, e.g., $\epsilon^1 = (1,0,0)$.

Let us now consider an element $\hat e_k$ of the Lie group $AN(3)$, which, as we will see in a moment, represents a group-valued momentum
\begin{equation}\label{II.1.19}
  \hat  e_k =e^{ik_i \hat x^i} e^{ik_0 \hat x^0}\,.
\end{equation}
In the representation (\ref{II.1.18}) this group element is  represented by  a $5\times 5$ matrix which
acts on $5$-dimensional Minkowski space as a linear transformation.
One finds
$$
\exp(i k_0\hat x^0)= \left(\begin{array}{ccc}
  \cosh\frac{k_0}\kappa & \mathbf{0} & \sinh\frac{k_0}\kappa \\&&\\
  \mathbf{0} & \mathbf{1} & \mathbf{0} \\&&\\
  \sinh\frac{k_0}\kappa\; & \mathbf{0}\; & \cosh\frac{k_0}\kappa
\end{array}\right)\, , \quad
\exp(i k_i\hat x^i) = \left(\begin{array}{ccc}
  1+\frac{\mathbf{k}^2}{2\kappa^2}\; &\; \frac{\mathbf k}\kappa \;& \; \frac{\mathbf{k}^2}{2\kappa^2}\\&&\\
  \frac{\mathbf k}\kappa & \mathbf{1} & \frac{\mathbf k}\kappa \\&&\\
  -\frac{\mathbf{k}^2}{2\kappa^2}\; & -\frac{\mathbf k}\kappa\; & 1-\frac{\mathbf{k}^2}{2\kappa^2}
\end{array}\right)\, ,
$$
where $\mathbf{1}$ is the unit $3\times 3$ matrix, and $\hat  e_k$ can be written in schematic form
\begin{equation}\label{II.1.19a}
 \hat  e_k   =\left(\begin{array}{ccc}
 \frac{ \bar p_4}\kappa \;&\;  \frac{\mathbf k}\kappa \; &\;
\frac{  p_0}\kappa\\&&\\
  \frac{\mathbf p}\kappa  & \mathbf{1} & \frac{\mathbf p}\kappa  \\&&\\
  \frac{\bar p_0}\kappa\; & -\frac{\mathbf k}\kappa\; &
\frac{  p_4}\kappa
\end{array}\right)\, ,
\end{equation}
where $p_0$, $p_i$ and $p_4$ are defined below, while $\bar{p}_0=\kappa  \sinh
\frac{k_0}{\kappa} - \frac{\mathbf{k}^2}{2\kappa}$, $\bar{p}_4=\kappa \cosh
\frac{k_0}{\kappa} + \frac{\mathbf{k}^2}{2\kappa}\, e^{
{k_0}/\kappa}$.

To describe the manifold of the group $AN(3)$ we choose a point in 5-dimensional Minkowski space, which becomes the momentum space origin
$\cal O$ with coordinates $(0,\ldots,0,\kappa)$ and act on it with the matrix $\hat  e_k$
(\ref{II.1.19a}), obtaining
$$
(p_0, {p}_i, p_4) = \hat  e_k \, \cal O
$$
On the left hand side we have coordinates of a point in the
$5$-dimensional Minkowski space, being in one to one correspondence with the group element $\hat  e_k$. The coordinates $(p_0, {p}_i, p_4)$
are related to the original parametrization $(k_0, k_i)$ of the group element as follows
\begin{eqnarray}
 {p_0}(k_0, \mathbf{k}) &=&\kappa  \sinh
\frac{k_0}{\kappa} + \frac{\mathbf{k}^2}{2\kappa}\,
e^{  {k_0}/\kappa}, \nonumber\\
 p_i(k_0, \mathbf{k}) &=&   k_i \, e^{  {k_0}/\kappa},\label{II.1.20}\\
 {p_4}(k_0, \mathbf{k}) &=& \kappa \cosh
\frac{k_0}{\kappa} - \frac{\mathbf{k}^2}{2\kappa}\, e^{
{k_0}/\kappa}.\nonumber
\end{eqnarray}
 There is a natural action of the $4$-dimensional Lorentz group on the $5$-dimensional Minkowski space, which takes the form
\begin{align}
\delta_\lambda p_0 &=\lambda^i \, p_i\,,\quad \delta_\lambda p_i =\lambda_i \, p_0\,,\quad \delta_\lambda p_4 =0\nonumber\\
\delta_\rho p_0 &=0\,,\quad \delta_\rho p_i =\epsilon_{ijk} \rho^j \, p^k\,,\quad \delta_\rho p_4 =0\nonumber
\end{align}
for infinitesimal boosts and rotations parameters $\lambda_i$, $\rho_i$. Since the Lorentian momenta components $p_0, \mathbf{p}$, transform as a vector,  $p_0^2-\mathbf{p}^2 $ is Lorentz-invariant and, as usual, the representations of the Lorentz group, in the spinless case that we consider here, are labelled by values of the mass $m^2$ and sign of energy $p_0$. Therefore the representations of the Poincar\'e algebra are characterized by mass-shell condition $p_0^2-\mathbf{p}^2=m^2$.

It is easy to check that\footnote{There are two solutions of the first equation in \eqref{II.1.21}, but since the point $\cal O$ for which $p_4 =1$ belongs to the solution we are interested in we choose $p_4$ positive.}
\begin{equation}\label{II.1.21}
   -p_0^2 + \mathbf{p}^2 + p_4^2 =\kappa^2\,, \quad p_4>0
\end{equation}
 It follows that the group $AN(3)$ is isomorphic, as a manifold, to a submanifold of the
$4$-dimensional de Sitter space. This submanifold is defined by the conditions
\begin{equation}\label{II.1.22}
    p_0+p_4 =\kappa e^{k_0/\kappa}>0\,,\quad p_4\equiv \sqrt{\kappa^2 +p_0^2 -\mathbf{p}^2}>0\,.
\end{equation}

On-shell $p_0^2-\mathbf{p}^2 = m^2$ and the condition \eqref{II.1.22} takes the form
\begin{equation}\label{0.2}
  p_0 +\sqrt{m^2+\kappa^2} >0
\end{equation}
Observe that this condition does not impose any restrictions on  positive energy states, but provides a lower bound on the negative energy ones $0>p_0> -\sqrt{m^2+\kappa^2} $. This condition seemed first to be Lorentz invariance violating \cite{Bruno:2001mw} because by acting with the Lorentz boost we can make $p_0$ acquire an arbitrary negative value, but was later shown to preserve Lorentz symmetry in a nontrivial way \cite{Arzano:2009ci}. To understand how it comes about let us introduce the antipodal map $S(p)$ defined as
\begin{equation}\label{0.4}
  S(p_0) = -p_0 + \frac{\mathbf{p}^2}{p_0+p_4} = \frac{\kappa^2}{p_0+p_4}-p_4\,,\quad S(\mathbf{p}) =-\frac{\kappa \mathbf{p} }{p_0+p_4}\,,\quad S(p_4) = p_4
\end{equation}
Notice that on-shell $S(\omega_p)=S(\sqrt{m^2+\mathbf{p}^2})$ is always negative.

It is worth mentioning in passing that if $p_0^2 - \mathbf{p}^2 =m^2 $ then $S(p_0)^2 -S(\mathbf{p})^2 = m^2$ and vice versa, so the former serves as an alternative form of mass-shell relation. As we will see both these mass shell conditions will arise in the theory of deformed scalar field.

One checks that this map provides a one-to-one correspondence between the `positive energy' submanifold $p_0>0$ and the negative energy one, satisfying the constraint \eqref{0.2}. Indeed take a positive energy state with energy $p_0>0$ and momentum $\mathbf{p}$ and apply the antipode to it. We find
$$
S(p_0)+p_4  = -p_0 + \frac{\mathbf{p}^2}{p_0+p_4} +p_4=\frac{\kappa^2}{p_0+p_4} >0
$$
We define the action of Lorentz symmetry on negative energy states by applying it to the corresponding positive energy one and taking the antipode of the result, schematically,
\begin{equation}\label{0.4b}
  L \triangleright S(p) \equiv S(L\triangleright p)\,,\quad p_0>0
\end{equation}
With this definition the orbits of Lorentz group for both positive and negative energies belong to the momentum space. We will describe the Lorentz transformations of the antipode in Appendix \ref{AppS}.

The coordinates $p_A$ \eqref{II.1.20} cover only   half of de Sitter momentum space.  It turns out (see below) that in order to construct a field with well defined properties under discrete spacetime symmetries, we have to introduce another, dual, momentum space defined as an orbit of $AN(3)$ group emanating from the point $\cal O^*$ with coordinates $(0,\ldots,0,-\kappa)$. These coordinates can be constructed with the help of a special element $\mathfrak{z}$ \cite{Freidel:2007hk} that maps $(0,\ldots,0,\kappa)$ to $(0,\ldots,0,-\kappa)$,
\begin{equation}\label{0.5a}
  \mathfrak{z}=  e^{\pi\kappa \hat X^0}= \left(\begin{array}{ccc}
  -1 & \mathbf{0} & 0 \\&&\\
  \mathbf{0} & \mathbf{1} & \mathbf{0} \\&&\\
  0\; & \mathbf{0}\; & -1
\end{array}\right)
\end{equation}
(or $\hat e_k$ in \eqref{II.1.19} with $k_i=0$, $k_0= -i\pi\kappa$).

We define
\begin{equation}\label{0.5b}
  \hat e^*_k=\hat e_k\,\mathfrak{z} = e^{ik_i \hat x^i} e^{ik_0\hat  x^0}\,\mathfrak{z}
\end{equation}
and acting with this group element on $(0,\ldots,0,\kappa)$, instead of \eqref{II.1.20} we get
\begin{eqnarray}
 {p_0}^*(k_0, \mathbf{k}) &=&-\kappa  \sinh
\frac{k_0}{\kappa} - \frac{\mathbf{k}^2}{2\kappa}\,
e^{  {k_0}/\kappa}, \nonumber\\
 p_i^*(k_0, \mathbf{k}) &=&  - k_i \, e^{  {k_0}/\kappa},\label{II.1.23}\\
 {p_4}^*(k_0, \mathbf{k}) &=& -\kappa \cosh
\frac{k_0}{\kappa} + \frac{\mathbf{k}^2}{2\kappa}\, e^{
{k_0}/\kappa}.\nonumber
\end{eqnarray}
with
\begin{equation}\label{II.1.24}
    p^*_0+p^*_4 = -\kappa e^{k_0/\kappa}<0\,,\quad p^*_4\equiv -\sqrt{\kappa^2 +(p^*_0)^2 -(\mathbf{p}^*)^2}<0\,.
\end{equation}
On-shell the condition \eqref{II.1.24} takes the form
\begin{equation}\label{0.2a}
  p^*_0 -\sqrt{m^2+\kappa^2} <0
\end{equation}
so that this time it does not impose any restrictions on  negative energy states, but provides an upper bound on the positive energy ones $0<p^*_0< \sqrt{m^2+\kappa^2} $. Again one solves the apparent problem with Lorentz symmetry with the help of the antipode, which has the form
\begin{equation}\label{0.6}
  S(p^*_0) =- p^*_0 + \frac{\mathbf{p}^*{}^2}{p^*_0+p^*_4} = \frac{\kappa^2}{p^*_0+p^*_4}-p_4^*\,,\quad S(\mathbf{p}^*) =\frac{\kappa \mathbf{p}^* }{p^*_0+p^*_4}\,,\quad S(p^*_4) = p^*_4
\end{equation}
On-shell $S(\omega^*_p) = S(-\sqrt{m^2+\mathbf{p}^*{}^2 })$ is always positive.

To formulate the field theory we must first describe the algebra of plane waves and differential calculus. We start with the group elements (also called `noncommutative' plane waves) $\hat e_k$ \eqref{II.1.19} (associated with the submanifold $p_0+p_4>0$) and $\hat e^*_k$ \eqref{0.5b} (for the submanifold $p_0+p_4<0$).  We use the five-dimensional Lorentz covariant differential calculus, see \cite{Freidel:2007hk} and references therein for details. To this end we introduce the spacetime derivatives $\hat\partial_\mu$ and an additional derivative in fourth direction $\hat\partial_4$ defined by their action on the plane waves
\begin{align}
\hat\partial_\mu\, \hat e_k &=i p_\mu(k)\, \hat e_k\,,\quad \hat\partial_4\, \hat e_k =i(\kappa- p_4(k))\, \hat e_k\nonumber\\
\hat\partial_\mu\, \hat e^*_k &= ip^*_\mu(k)\, \hat e^*_k\,,\quad \hat\partial_4\, \hat e^*_k =i(\kappa- p^*_4(k))\, \hat e^*_k\label{0.7}
\end{align}

Following~\cite{Freidel:2007hk} we define the Weyl map\footnote{ Notice that the choice of Weyl map is not unique (see for instance~\cite{AmelinoCamelia:2001fd} for a different choice, and the discussion in~\cite{Arzano:2018gii}), and from this choice depend also the star product structures. In this paper we choose to adopt the Weyl map introduced in~\cite{Freidel:2007hk}, mapping "time-to-the-right" ordered non-commutative plane waves to standard exponentials of commutative coordinates, expressed in terms of “embedding” momenta $p_{A}\left(k\right)$ ($A=0,1,\dots,4$).} $\cal W$ that maps group elements (plane waves on non-commutative $\kappa$-Minkowski spacetime) to ordinary plane waves on commutative spacetime manifold with coordinates $x$, as
\begin{equation}\label{0.8}
  {\cal W}(\hat e_k(\hat x)) = e_p(x)
\end{equation}
defined by the action of the derivatives
\begin{equation}\label{0.9}
  {\cal W}(\hat\partial_\mu \hat e_k)(\hat x) = \partial_\mu e_p(x)\,,\quad{\cal W}(\hat\partial_\mu \hat e^*_k)(\hat x) = \partial_\mu e^*_p(x)
\end{equation}
with $\partial_\mu$ being the standard partial derivative\footnote{An explicit realization of this star product was presented in \cite{KowalskiGlikman:2009zu}}. The star product presented here  coincides with the one proposed in \cite{Mercati:2011pv} and further discussed in \cite{Matassa:2013gva}, \cite{Poulain:2018mcm} and \cite{Mathieu:2020ccc}. It follows that
\begin{equation}\label{0.10}
  e_p(x) = e^{ip_\mu\, x^\mu} = e^{-i(\omega_\mathbf{p}t-\mathbf{p}\mathbf{x})}\,,\quad e^*_p(x) = e^{ip^*_\mu\, x^\mu} = e^{-i(\omega_\mathbf{p}^*t-\mathbf{p}^*\mathbf{x})}
\end{equation}
with the on-shell relations
\begin{equation}\label{0.11}
  \omega_\mathbf{p}= \sqrt{m^2 +p^2}\,,\quad \omega^*_\mathbf{p} =-\sqrt{m^2 +p^*{}^2}\,,\quad p_4= \sqrt{m^2 +\kappa^2}\,,\quad p_4^* =-\sqrt{m^2 +\kappa^2}
\end{equation}

The Weyl map makes it possible to construct the star product of two commuting plane waves from the product of two group elements
\begin{equation}\label{0.12}
  {\cal W}(\hat e_k\,\hat e_l) \equiv e_{p(k)}\star e_{q(l)} = e_{p\oplus q}
\end{equation}
In the case of two positive energy plane waves  we have
\begin{equation}\label{0.13}
  \hat e_k\,\hat e_l = \hat e_{k\oplus l}
\end{equation}
with
\begin{equation}\label{0.14}
  (k\oplus l)_0 = k_0 + l_0\,,\quad  (k\oplus l)_i = k_i + e^{-k_0/\kappa}\, l_i
\end{equation}
Then acting with the group element \eqref{0.13} on the reference vector $(0,\ldots,0,\kappa)$ we get
\begin{align}
(p\oplus q)_0 &= \frac1\kappa\, p_0(q_0+q_4) + \frac{\mathbf{p}\mathbf{q}}{p_0+p_4} +\frac{\kappa}{p_0+p_4}\, q_0\nonumber\\
(p\oplus q)_i &=\frac1\kappa\, p_i(q_0+q_4) + q_i\nonumber\\
(p\oplus q)_4 &= \frac1\kappa\, p_4(q_0+q_4) - \frac{\mathbf{p}\mathbf{q}}{p_0+p_4} -\frac{\kappa}{p_0+p_4}\, q_0\label{0.15}
\end{align}

Let us use the same construction in the case of the negative energy plane waves. To this end we must first compute the product
\begin{equation}\label{0.16}
\mathfrak{z}\, e_{(p_0 , \mathbf{p})}=e_{(p_0 , -\mathbf{p})}\mathfrak{z}\,.
\end{equation}
From
\begin{equation}\label{0.16a}
  {\cal W}(\hat e^*_k\,\hat e_l) \equiv e^*_{p(k)}\star e_{q(l)} = e_{p^*\oplus q}
\end{equation}
we find
\begin{align}
(p^*\oplus q)_0 &= \frac1\kappa\, p^*_0(q_0+q_4) + \frac{\mathbf{p}^*\mathbf{q}}{p^*_0+p^*_4} +\frac{\kappa}{p^*_0+p^*_4}\, q_0\nonumber\\
(p^*\oplus q)_i &=\frac1\kappa\, p^*_i(q_0+q_4) + q_i\nonumber\\
(p^*\oplus q)_4 &= \frac1\kappa\, p^*_4(q_0+q_4) - \frac{\mathbf{p}^*\mathbf{q}}{p^*_0+p^*_4} -\frac{\kappa}{p^*_0+p^*_4}\, q_0\label{0.17}
\end{align}
(To compute this, one starts with \eqref{0.15}, changes the overall sign, then changes the sign of $p$ replacing it by ${p}^*$, and finally changes the sign of $\mathbf{q}$ according to \eqref{0.16}.)

Similarly
\begin{align}
(p\oplus q^*)_0 &= \frac1\kappa\, p_0(q^*_0+q^*_4) + \frac{\mathbf{p}\mathbf{q}^*}{p_0+p_4} +\frac{\kappa}{p_0+p_4}\, q^*_0\nonumber\\
(p\oplus q^*)_i &=\frac1\kappa\, p_i(q^*_0+q^*_4) + q^*_i\nonumber\\
(p\oplus q^*)_4 &= \frac1\kappa\, p_4(q^*_0+q^*_4) - \frac{\mathbf{p}\mathbf{q}^*}{p_0+p_4} -\frac{\kappa}{p_0+p_4}\, q^*_0\label{0.18}\,.
\end{align}

Finally, we consider the composition of two negative energy plane waves. (In this case after moving through the $Q$ plane wave we get $\mathfrak{z}^2=1$)
\begin{align}
(p^*\oplus q^*)_0 &= \frac1\kappa\, p^*_0(q^*_0+q^*_4) + \frac{\mathbf{p}^*\mathbf{q}^*}{p^*_0+p^*_4} +\frac{\kappa}{p^*_0+p^*_4}\, q^*_0\nonumber\\
(p^*\oplus q^*)_i &=\frac1\kappa\, p^*_i(q^*_0+q^*_4) + q^*_i\nonumber\\
(p^*\oplus q^*)_4 &= \frac1\kappa\, p^*_4(q^*_0+q^*_4) - \frac{\mathbf{p}\mathbf{q}^*}{p^*_0+p^*_4} -\frac{\kappa}{p^*_0+p^*_4}\, q^*_0\label{0.19}\,.
\end{align}
Notice that, remarkably, all the composition laws \eqref{0.15}-\eqref{0.19} have exactly the same form so there is no need to distinguish between them.

 Let us finish this section with the definition of an adjoint of the plane wave. For the noncommutative plane wave $\hat e_k$ its adjoint  $\hat e^\dag_k$ is defined by the condition
\begin{equation}\label{adjo1}
  \hat e_k \hat e^\dag_k = \hat e^\dag_k \hat e_k  = 1
\end{equation}
from which it follows that
\begin{equation}\label{adjo2}
  \hat e^\dag_k = \hat e_{S(k)}
\end{equation}
Accordingly, in the star product formalism we express these equations as
\begin{equation}\label{adjo3}
   e_p\star  e^\dag_q =  e^\dag_q \star e_p  = 1
\end{equation}
from which it follows that
\begin{equation}\label{adjo4}
   e^\dag_p =  e_{S(p)}
\end{equation}
The analogous expressions for $p^*_A$ coordinates are easy to obtain.

\section{Action and field equations}

 Having discussed all the necessary technical tools in the preceding section we can now turn to the construction of the theory of free complex scalar field. As customary in non-commutative field theories, we define a notion of integral on non-commutative space-time via the Weyl (or quantization) map~(\ref{0.8}). In particular we set
\begin{align}
\widehat \int \hat e_k(\hat x) := \int_{\mathbb{R}^4} d^4 x \mathcal{W}(\hat e_k(\hat x)) = \int_{\mathbb{R}^4} d^4 x \, e^{i p x}.
\label{integration}
\end{align}
Fields on $\kappa$-Minkowski can be defined in terms of a suitable ``noncommutative'' (or, for some authors, quantum group-) Fourier transform~\cite{Freidel:2007hk,Arzano:2018gii,Freidel:2005bb,Freidel:2005ec,Guedes:2013vi,Oriti:2018bwr}. In accordance with our choice of Weyl map, we adopt the noncommutative Fourier transform introduced in~\cite{Freidel:2007hk}:
\begin{align}
\hat \phi (\hat x) = \int_{AN(3)} d \mu (p) \tilde \phi(p) \hat e_k(\hat x)
\end{align}
and its inverse
\begin{align}
\tilde \phi (p) = \widehat \int \hat e^{\dagger}_k(\hat x) \phi(\hat x),
\end{align}
where the measure $d \mu (p)$ is the $AN(3)$ left-invariant measure
\begin{equation}\label{measure4d}
d \mu (p) = \frac{d^4p}{p_4/\kappa}\Big|_{p_+>0 \ \& \ p_4 = \sqrt{\kappa^2 + p_0^2 - \mathbf{p}^2}} \ ,
\end{equation}
and the coordinates $p$ are intended as the "embedding" coordinates $p(k)$ given by~(\ref{II.1.20}).
The definition can be thus extended to fields of commutative coordinates through Weyl map
\begin{equation}\label{phix}
\phi(x) := {\cal W}(\hat \phi(\hat x)).
\end{equation}
Explicitly
\begin{equation}\label{phixExpl}
\phi(x) = \int_{AN(3)} d \mu (p) \tilde \phi(p) e_p(x).
\end{equation}
Notice that the $\phi(x)$ defined by~(\ref{phix}) and~(\ref{phixExpl}) depend on the choice of Weyl map. In the explicit expression~(\ref{phixExpl}), the dependence is encoded in both the measure of integration, expressed in terms of embedding momenta $p$ restricted to the $AN(3)$ manifold, and on the Fourier "coefficients" $\tilde{\phi}(p)$.

From~(\ref{0.12}) and~(\ref{integration}) it follows that the inverse noncommutative Fourier transform can be expressed as
\begin{align}
\tilde \phi (p) = \int_{\mathbb{R}^4} e^{\dagger}_p( x) \star \phi( x),
\end{align}
and that the noncommutative product extends to a star product of fields of commutative coordinates
\begin{align}
{\cal W} (\hat \phi (\hat x) \hat \psi (\hat x) ) = \phi(x) \star \psi(x).
\end{align}
In particular we have the following useful identity
\begin{align}
\widehat \int \hat \phi (\hat x) \hat \psi (\hat x) = \int_{\mathbb{R}^4} \phi(x) \star \psi(x).
\end{align}
The star product here coincides with the one defined in \cite{Mercati:2011pv} (generalized to 4d), which can be checked by calculating that it gives the identical result for the coordinate functions $x^{\mu}$. However it is not clear if the construction of the integral/twisted trace presented in that paper coincides with our definition of the integral.

Using the non-commutative Fourier transform and the star-product, we can formulate the action of free fields on $\kappa$-Minkowski space-time as a standard integral action in terms of (properly defined as above) fields of commutative coordinates. In particular, we  define the action to be an integral of the bilinear hermitian expression, in fields and derivatives, obtained with the help of the star product.  The integral satisfies the exchange properties for the plane waves \cite{Freidel:2007hk}
\begin{equation}\label{4.8}
  \int_{ \mathbb{R}^4}d^4x\,  e^\dag_p\star   e_q=  \int_{ \mathbb{R}^4}d^4x\,  e^\dag_q \star  e_p
\end{equation}
and the most general expression for the hermitian action is
\begin{equation}\label{4.9}
  S = \frac{1}{2} \int_{ \mathbb{R}^4}d^4x\,
  \left[ (\partial_\mu \phi)^\dag\star\partial^\mu \phi
  + (\partial_\mu \phi) \star(\partial^\mu \phi)^\dag
  - m^2 (\phi^\dag\star \phi + \phi \star \phi^\dag)\right]
\end{equation}
In order to compute the variation of the action and to derive field equations, we have to make use of the $\star$-integration by parts, which is described in detail in Appendix \ref{Intbyparts}.
 Writing $S = \frac{1}{2}(S_1 + S_2)$ where
\begin{align}\label{4.34-3}
	S_1
	&=
	\int_{\mathbb{R}^4} d^4x \,\, (\partial^\mu \phi)^\dag \star (\partial_\mu \phi) - m^2 \phi^\dag \star \phi
\end{align}
and
\begin{align}\label{4.34-4}
	S_2
	&=
	\int_{\mathbb{R}^4} d^4x \,\ (\partial_\mu \phi) \star (\partial^\mu \phi)^\dag  - m^2 \phi \star \phi^\dag .
\end{align}
we find
\begin{align}\label{4.35}
	\delta S_1 = \frac12\int_{ \mathbb{R}^4}d^4x\,
	(\partial_\mu \delta \phi)^\dag\star\partial^\mu \phi
	+
	(\partial_\mu \phi)^\dag\star\partial^\mu \delta \phi
	-
	m^2 \delta \phi^\dag\star \phi
	-
	m^2 \phi^\dag\star \delta \phi
\end{align}
which can be rewritten as
\begin{align}\label{4.36}
	\delta S_1 = \frac12\int_{ \mathbb{R}^4}d^4x\,
	\Bigg\{
	-&\frac{\Delta_+}{\kappa}
	\left[
	(\partial_\mu^\dag (\partial^\mu)^\dag - m^2) \phi^\dag
	\star \delta\phi
	\right]
	+ \partial_A \left( \Pi^A \star \delta \phi \right) \nonumber \\
	-&
	\frac{\kappa}{\Delta_+}
	\left[
	\delta\phi^\dag
	\star
	(\partial_\mu\partial^\mu - m^2) \phi
	\right]
	+
	\partial_A^\dag
	\left(
	\delta\phi^\dag \star \left(\Pi^A\right)^\dag
	\right)
	\Bigg\}
\end{align}
where
\begin{align}
	\Pi^0_1 = (\Pi_0)_1 &= \frac{1}{\kappa} (\Delta_+ \partial_0^\dag +i m^2) \phi^\dag  \label{4.37} \\
	\Pi^i_1 = -(\Pi_i)_1 &= (-\partial_i (1 + i\Delta_+^{-1}\partial_0))\phi^\dag  \label{4.38} \\
	\Pi^4_1=(\Pi_4)_1 &= -i\frac{m^2\phi^\dag}{\kappa} \label{4.39}
\end{align}
and, analogously,
\begin{align}\label{4.40}
	\delta S_2 = \frac12\int_{ \mathbb{R}^4}d^4x\,
	\partial^\mu \phi \star (\partial_\mu \delta \phi)^\dag
	+
	\partial^\mu \delta \phi \star (\partial_\mu \phi)^\dag
	-
	m^2  \phi \star \delta\phi^\dag
	-
	m^2 \delta \phi \star \phi^\dag
\end{align}
which can be rewritten as
\begin{align}\label{4.41}
	\delta S_2 = \frac12\int_{ \mathbb{R}^4}d^4x\,
	\Bigg\{
	-&
	\left[
	\delta\phi \star
	(\partial_\mu^\dag (\partial^\mu)^\dag - m^2) \phi^\dag
	\right]
	+ \partial_A \left( \delta \phi \star \Pi^A  \right) \nonumber \\
	-&
	\left[
	(\partial_\mu\partial^\mu - m^2) \phi
	\star
	\delta\phi^\dag
	\right]
	+
	\partial_A^\dag
	\left(
	\left(\Pi^A\right)^\dag
	\star
	\delta\phi^\dag
	\right)
	\Bigg\}
\end{align}
where
\begin{align}
	\Pi^0_2 = (\Pi_0)_2 &= \left(\frac{\kappa}{\Delta_+} \partial_0^\dag +\frac{i}{\kappa}(\partial_0^\dag)^2
	\right)\phi^\dag  \label{4.42} \\
	\Pi^i_2 = -(\Pi_i)_2 &= - \frac{\kappa}{\Delta_+} (\partial_i^\dag + i\partial_i\partial_0^\dag)\phi^\dag \label{4.43}  \\
	\Pi^4_2=(\Pi_4)_2 &= +i\frac{(\partial_0^\dag)^2}{\kappa} \phi^\dag . \label{4.44}
\end{align}
Therefore the field equations have the form
\begin{equation}\label{Feqs}
  (\partial_\mu\partial^\mu - m^2) \phi =0\,,\quad (\partial_\mu^\dag (\partial^\mu)^\dag - m^2) \phi^\dag =0
\end{equation}
which, as we will see below, lead to two non-trivially related mass-shell conditions, describing the same orbit of the Lorentz group on the momentum manifold.

\section{The complex scalar field}\label{CSF}
 Now we are in position to formulate the theory of the
deformed free complex scalar field. In what follows we will use the
strategy adopted in \cite{Freidel:2007hk} of developing the non-commutative
field theory in terms of fields on commutative Minkowski space-time
equipped with a non-commutative star-product. Using the identity (cf.\ \eqref{adjo1}--\eqref{adjo4})
\[
e^{-ipx}\star e^{-iS(p)x}=e^{-iS(p)x}\star e^{-ipx}=1
\]
to define the adjoint of the plane wave
\begin{equation}
\left(e^{-ipx}\right)^{\dag}=e^{-iS(p)x},\label{1.3}
\end{equation}
we can write the adjoint field as
\begin{equation}
\phi^{\dagger}\left(\hat{x}\right)=\int d\mu\left(p\left(k\right)\right)\,\tilde{\phi}^{\dagger}\left(p\right){\cal W}\left(e^{-iS\left(p\left(k\right)\right)x}\right),
\end{equation}
and one can define
\begin{equation}
\phi^{\dagger}\left(x\right)={\cal W}^{-1}\left(\phi^{\dagger}\left(\hat{x}\right)\right)=\int d\mu\left(p\right)\,\tilde{\phi}^{\dagger}\left(p\right)e^{-iS\left(p\right)x}.
\end{equation}
Changing integration variables in the last expression, and using that
$S\left(S\left(p\right)\right)=p$, we can rewrite it as
\begin{equation}
\phi^{\dagger}\left(x\right)=\kappa^3\int d\mu\left(p\right)\,p_{+}^{-3}\tilde{\phi}^{\dagger}\left(S\left(p\right)\right)e^{-ipx},\label{complexPhi}
\end{equation}
where we used\footnote{Notice in passing that the r.h.s. of (\ref{measureAntipode}) coincides
with the right invariant on $AN_{3}$, as one can check from the multiplication
of two group elements. If we denote the left invariant measure
we are using as $d\mu_{L}\left(p\right)$, one thus have the property
that under antipode, $d\mu_{L}\left(S\left(p\right)\right)=d\mu_{R}\left(p\right)$.
This property is indeed a manifestation of the fact that the antipode
map on the manifold corresponds to the inversion on the group elements.}
\begin{equation}
d\mu\left(S\left(p\right)\right)=\frac{\kappa^3}{p_{+}^3}d\mu\left(p\right),\label{measureAntipode}
\end{equation}
as one can easily check.

It follows, by comparing (\ref{complexPhi}) with (\ref{phixExpl}),
that the condition for $\phi\left(x\right)$ to be real is~\footnote{The same result was obtained in~\cite{Daszkiewicz:2004xy} working with the $k$ parametrization.}
\begin{equation}
\tilde{\phi}^{\dagger}\left(p\right)=\kappa^{-3}S^{3}\left(p_{+}\right)\tilde{\phi}\left(S\left(p\right)\right),
\end{equation}
or equivalently
\begin{equation}
\tilde{\phi}^{\dagger}\left(S\left(p\right)\right)=\kappa^{-3}p_{+}^{3}\tilde{\phi}\left(p\right),\label{realFieldCondition}
\end{equation}
where we considered that $S\left(p_{+}\right)=\kappa^2 p_{+}^{-1}$. We will discuss real fields in the forthcoming paper and here we will concentrate on the complex fields only.

According to the properties of the momentum space manifold described
in Sec.~\ref{sec:preliminaries} (see especially Eq.~(\ref{II.1.22})), the left-invariant
Haar measure on $AN\left(3\right)$ can be rewritten as the ordinary
Lebesgue measure on a restricted 5-dimensional momentum space with
(the factor $2\kappa$ here is included is for dimensional reasons)
\begin{equation}
d\mu\left(p\right)=2\kappa d^{5}p\ \delta\left(p_{0}^{2}-{\bf p}^{2}-p_{4}^{2}+\kappa^{2}\right)\theta\left(p_{+}\right)\theta\left(p_{4}\right).
\end{equation}
Let us now consider a field on the mass shell defined by $m$, that
we can write as ($A=0,1,\dots,4$)
\begin{equation}
\phi\left(x\right)=\int d^{5}p\ 2\kappa\delta\left(p_{A}p^{A}+\kappa^{2}\right)\theta\left(p_{+}\right)\theta\left(p_{4}\right)\delta\left(p_{\mu}p^{\mu}-m^{2}\right)\tilde{\phi}\left(p\right)e^{-ipx}
\end{equation}
One way of splitting the $\delta\left(p_{\mu}p^{\mu}-m^{2}\right)$
into ``positive and negative energy'' solutions, is to rewrite it
as
\begin{equation}
\label{splitOnshell}
\delta\left(p_{\mu}p^{\mu}-m^{2}\right)=\delta\left(p_{\mu}p^{\mu}-m^{2}\right)\theta\left(p_{0}-m\right)+\delta\left(p_{\mu}p^{\mu}-m^{2}\right)\theta\left(-p_{0}-m\right).
\end{equation}
Using this, we can rewrite the field as
\begin{equation}
\begin{split}\phi\left(x\right)= & \phi_{+}\left(x\right)+\phi_{-}\left(x\right)\\
= & \int d^{5}p\ 2\kappa\delta\left(p_{A}p^{A}+\kappa^{2}\right)\theta\left(p_{+}\right)\theta\left(p_{4}\right)\delta\left(p_{\mu}p^{\mu}-m^{2}\right)\theta\left(p_{0}-m\right)\tilde{\phi}\left(p\right)e^{-ipx}\\
 & +\int d^{5}p\ 2\kappa\delta\left(p_{A}p^{A}+\kappa^{2}\right)\theta\left(p_{+}\right)\theta\left(p_{4}\right)\delta\left(p_{\mu}p^{\mu}-m^{2}\right)\theta\left(-p_{0}-m\right)\tilde{\phi}\left(p\right)e^{-ipx},
\end{split}
\label{phiPlusMinus}
\end{equation}
where $\phi_{+}\left(x\right)$ and $\phi_{-}\left(x\right)$ denote
the ``positive and negative energy'' components of the onshell field.
Consider the ``negative energy'' part $\phi_{-}\left(x\right)$.
From the properties of the antipode map
\begin{equation}
\begin{gathered}S\left(p_{\mu}\right)S\left(p^{\mu}\right)=p_{\mu}p^{\mu},\\
S\left(p_{4}\right)=p_{4},
\end{gathered}
\end{equation}
that imply also $S\left(p_{A}\right)S(p^{A})=p_{A}p^{A}$, if we change
the integration variables as $p\rightarrow S\left(p\right)$, and
use that $S\left(S\left(p\right)\right)=p$ and (\ref{measureAntipode}),
we can rewrite $\phi_{-}\left(x\right)$ as
\begin{equation}
\begin{split}\phi_{-}\left(x\right)= & \int d^{5}S\left(p\right)\ 2\kappa\delta\left(S\left(p_{A}\right)S(p^{A})+\kappa^{2}\right)\theta\left(S\left(p_{+}\right)\right)\theta\left(S\left(p_{4}\right)\right)\\
 & \times\,\delta\left(S\left(p_{\mu}\right)S\left(p^{\mu}\right)-m^{2}\right)\theta\left(-S\left(p_{0}\right)-m\right)\tilde{\phi}\left(S\left(p\right)\right)e^{-iS\left(p\right)x}\\
= & \int d^{5}p\ 2\kappa\delta\left(p_{A}p^{A}+\kappa^{2}\right)\theta\left(p_{+}\right)\theta\left(p_{4}\right)\\
 & \times\delta\left(p_{\mu}p^{\mu}-m^{2}\right)\theta\left(-S\left(p_{0}\right)-m\right)S\left(p_{+}^{3}\right)\tilde{\phi}\left(S\left(p\right)\right)e^{-iS\left(p\right)x}\,,
\end{split}
\end{equation}
where we take into account the property
\begin{equation}
\theta\left(S\left(p_{+}\right)\right)=\theta\left(p_{+}^{-1}\right)=\theta\left(p_{+}\right).
\end{equation}
Now, notice that (accordingly to the discussion of Sec.~\ref{sec:preliminaries}),
\begin{equation}
\text{if}\quad p_{4}>0\quad\&\quad p_{+}>0\quad\&\quad p_{\mu}p^{\mu}=m^{2},\quad\Rightarrow\quad S\left(p\right)_{0}<-m\quad\Leftrightarrow\quad p_{0}>m,
\end{equation}
The proof is straightforward, since, on the mass shell,
\begin{equation}
S\left(p\right)_{0}=\frac{-p_{0}^{2}+{\bf p}^{2}-p_{0}p_{4}}{p_{+}}=\frac{-m^{2}-p_{0}p_{4}}{p_{+}},
\end{equation}
thus,
\begin{equation}
\begin{split}S\left(p\right)_{0}<-m\quad\Rightarrow\quad & -m^{2}-p_{0}p_{4}<-mp_{+}=-m\left(p_{0}+p_{4}\right)\\
\Rightarrow\quad & p_{0}>m.
\end{split}
\end{equation}
The proof that $p_{0}>m$ implies $S\left(p\right)_{0}<-m$,
is also straightforward. This shows that, for $p_{+}>0$
and $p_{4}>0$, i.e. on the $AN(3)$ submanifold we are interested
in (i.e. on that section of the de Sitter hyperboloid selected by
the measure $d\mu\left(p\right)$) the antipode acts indeed as a bijective
map that splits the positive and negative energy parts of the manifold
belonging to the same mass shell, as argued in Sec.~\ref{sec:preliminaries}, and in agreement with the observations reported in~\cite{Arzano:2009ci}. Since
the map is bijective (one-to-one), we can then interchange the $\theta\left(-S\left(p\right)_{0}-m\right)$
with the $\theta\left(p_{0}-m\right)$ in the integral, and rewrite
finally $\phi_{-}\left(x\right)$ as
\begin{equation}
\begin{split}\phi_{-}\left(x\right)= & \kappa^{-3}\int d^{5}p\ 2\kappa\delta\left(p_{A}p^{A}+\kappa^{2}\right)\theta\left(p_{+}\right)\theta\left(p_{4}\right)\\
 & \times\delta\left(p_{\mu}p^{\mu}-m^{2}\right)\theta\left(p_{0}-m\right)S\left(p_{+}^{3}\right)\tilde{\phi}\left(S\left(p\right)\right)e^{-iS\left(p\right)x}\ .
\end{split}
\label{phiMinus}
\end{equation}

If the field is real, condition (\ref{realFieldCondition}) holds,
and we have obtained the following result: on the $AN(3)$ measure
the on-shellness condition naturally splits the field into positive
and negative energy components, that are conjugate with each other,
with the antipode playing the role of conjugation for the plane wave,
i.e.
\begin{equation}
\begin{split}\phi\left(x\right)= & \int d\mu\left(p\right)\ \delta\left(p_{\mu}p^{\mu}-m^{2}\right)\theta\left(p_{0}-m\right)\left[\tilde{\phi}\left(p\right)e^{-ipx}+\tilde{\phi}^{\dagger}\left(p\right)e^{-iS\left(p\right)x}\right]\\
= & \int\frac{d^{3}p}{2\omega_{{\bf p}}\,p_{4}/\kappa}\left[\tilde{\phi}\left(\omega_{{\bf p}},{\bf p}\right)e^{-i\left(\omega_{{\bf p}}t-{\bf p}\cdot{\bf x}\right)}+\tilde{\phi}^{\dagger}\left(\omega_{{\bf p}},{\bf p}\right)e^{-i\left(S\left(\omega_{{\bf p}}\right)t-S\left({\bf p}\right)\cdot{\bf x}\right)}\right],
\end{split}
\end{equation}
where in the last row $p_{4}$ is ``onshell'', $p_{4}=\sqrt{m^{2}+\kappa^{2}}$.

For a complex field, it will be convenient to define the antiparticle
states, i.e. the ones associated to the negative energy part of the
field, as the ones associated to the dual (starred) copy of momentum
space. We first substitute, for $\phi_{-}\left(x\right)$, $p\rightarrow-p=p^{*}$,
so that (since $S\left(p^{*}\right)=S\left(-p\right)=-S\left(p\right)$),
it becomes
\begin{equation}
\begin{split}\phi_{-}\left(x\right)= & \kappa^{-3}\int d^{5}p^{*}\ 2\kappa\delta\left(p_{A}^{*}p_{*}^{A}+\kappa^{2}\right)\theta\left(-p_{+}^{*}\right)\theta\left(-p_{4}^{*}\right)\\
 & \times\delta\left(p_{\mu}^{*}p_{*}^{\mu}-m^{2}\right)\theta\left(-p_{0}^{*}-m\right)\left[-S\left(p_{+}^{*3}\right)\tilde{\phi}\left(-S\left(p^{*}\right)\right)\right]e^{iS\left(p^{*}\right)x}\\
= & \kappa^{-3}\int\frac{d^{3}p}{2|\omega_{{\bf p}}^{*}|\,p_{4}^{*}/\kappa}S(p_{+}^{*3})\tilde{\phi}\left(-S(\omega_{{\bf p}}^{*}),-S({\bf p}^{*})\right)e^{i\left(S(\omega_{{\bf p}}^{*})t-S({\bf p}^{*})\cdot{\bf x}\right)}\ ,
\end{split}
\end{equation}
where  $p_{4}^{*}=-\sqrt{m^{2}+\kappa^{2}}$. Thus, using
(\ref{phiPlusMinus}), we have the expansion
\begin{equation}
\begin{split}\phi\left(x\right)= & \int\frac{d^{3}p}{2\omega_{{\bf p}}\,p_{4}/\kappa}\tilde{\phi}\left(\omega_{{\bf p}},{\bf p}\right)e^{-i\left(\omega_{{\bf p}}t-{\bf p}\cdot{\bf x}\right)}\\
 & + \kappa^{-3}\int\frac{d^{3}p}{2|\omega_{{\bf p}}^{*}|\,p_{4}^{*}/\kappa}S(p_{+}^{*3})\tilde{\phi}\left(-S(\omega_{{\bf p}}^{*}),-S({\bf p}^{*})\right)e^{i\left(S(\omega_{{\bf p}}^{*})t-S({\bf p}^{*})\cdot{\bf x}\right)}\ .
\end{split}
\end{equation}

Since the mass-shell condition
\begin{equation}
p_{0}^{2}-\mathbf{p}^{2}=m^{2}\,\quad\mbox{or}\quad S(p_{0})^{2}-S(\mathbf{p})^{2}=m^{2}\label{1.1}
\end{equation}
has the standard classical form, we would like to define the Fourier
components of the complex field as close as possible as the classical
expression \cite{Weinberg:1995mt} in terms of creation and annihilation
operators
\begin{equation}
\phi_{+}(x)\sim\int\frac{d^{3}p}{\sqrt{2\omega_{{\bf p}}}}\,a_{\mathbf{p}}\,e^{-i(\omega_{p}t-\mathbf{p}\mathbf{x})},\qquad\phi_{-}(x)\sim\int\frac{d^{3}p}{\sqrt{2\omega_{{\bf p}}}}\,b_{\mathbf{p}}^{\dagger}\,e^{i(\omega_{p}t-\mathbf{p}\mathbf{x})}\ .\label{1.2}
\end{equation}
We postulate
\begin{equation}
a_{{\bf p}}=\frac{\xi^{-1}\left(p\right)}{\sqrt{2\omega_{{\bf p}}}\,p_{4}/\kappa}\tilde{\phi}\left(\omega_{{\bf p}},{\bf p}\right),\qquad b_{{\bf p}^{*}}= \kappa^{-2} \frac{\xi^{-1}(p^{*})}{\sqrt{2|\omega_{{\bf p}}|}\,p_{4}^{*}}S(p_{+}^{*3})\tilde{\phi}^{\dagger}\left(-S\left(\omega_{{\bf p}}^{*}\right),-S\left({\bf p}^{*}\right)\right)\ ,
\end{equation}
where we include an additional factor ($p_{+}$ has to be considered
onshell, $p_{+}=\sqrt{{\bf p}^{2}+m^{2}}+\sqrt{\kappa^{2}+m^{2}}$)
\begin{equation}
\label{xi}
\xi\left(p\right)=\left(1+\frac{|p_{+}|^{3}}{\kappa^{3}}\right)^{-\frac{1}{2}},
\end{equation}
that makes the form of the momentum space action, which we will make
use of later, particularly simple.

Finally, we have, for the on-shell complex field and its adjoint, the
expansions
\begin{align}
\phi(x) & =\int\frac{d^{3}p}{\sqrt{2\omega_{p}}}\,\left[1+\frac{|p_{+}|^{3}}{\kappa^{3}}\right]^{-\frac{1}{2}}a_{\mathbf{p}}\,e^{-i(\omega_{p}t-\mathbf{p}\mathbf{x})}+\int\frac{d^{3}p^{*}}{\sqrt{2|\omega_{p}^{*}|}}\,\left[1+\frac{|p_{+}^{*}|^{3}}{\kappa^{3}}\right]^{-\frac{1}{2}}b_{\mathbf{p^{*}}}^{\dag}\,e^{i(S(\omega_{p}^{*})t-S(\mathbf{p}^{*})\mathbf{x})}\nn\label{fa}\\
 & \equiv\phi_{(+)}(x)+\phi_{(-)}(x)
\end{align}
\begin{align}
\phi^{\dag}(x) & =\int\frac{d^{3}p}{\sqrt{2\omega_{p}}}\,\left[1+\frac{|p_{+}|^{3}}{\kappa^{3}}\right]^{-\frac{1}{2}}a_{\mathbf{p}}^{\dag}\,e^{-i(S(\omega_{p})t-S(\mathbf{p})\mathbf{x})}+\int\frac{d^{3}p^{*}}{\sqrt{2|\omega_{p}^{*}|}}\,\left[1+\frac{|p_{+}^{*}|^{3}}{\kappa^{3}}\right]^{-\frac{1}{2}}b_{\mathbf{p^{*}}}\,e^{i(\omega_{p}^{*}t-\mathbf{p}^{*}\mathbf{x})}\nn\label{fb}\\
 & \equiv\phi_{(+)}^{\dag}(x)+\phi_{(-)}^{\dag}(x)
\end{align}
Since $\omega_{p}>0$ and $S(\omega_{p}^{*})>0$ the field \eqref{fa}
is a combination of positive energy particle states and negative energy
antiparticle ones, while in \eqref{fa} we have the opposite arrangement,
as it should be. This particular definition of the field and its adjoint,
contrary to earlier approaches where to define the field and its adjoint
only one portion of de Sitter space was used, allows for simple action
of discrete symmetries, see Section \ref{CPT} below for the details.

From eq. \eqref{4.36} one sees that the equations of motion (EOM)
for the field $\phi$ are indeed the expected ones. Furthermore, one
can get the EOM also for the $a_{\mathbf{p}}$, $a_{\mathbf{p}}^{\dag}$,
$b_{\mathbf{p^{*}}}$ and $b_{\mathbf{p^{*}}}^{\dag}$ by applying
the EOM to the fields in eq. \eqref{fa}, \eqref{fb}. We get
\begin{align}
(\partial_{\mu}\partial^{\mu}-m^{2})\phi= & \int\frac{d^{3}p}{\sqrt{2\omega_{p}}}\left[1+\frac{|p_{+}|^{3}}{\kappa^{3}}\right]^{-\frac{1}{2}}\,(p_{\mu}p^{\mu}-m^{2})a_{\mathbf{p}}\,e^{-i(\omega_{p}t-\mathbf{p}\mathbf{x})}\label{EOMa}\\
+ & \int\frac{d^{3}p^{*}}{\sqrt{2|\omega_{p}^{*}|}}\left[1+\frac{|p_{+}|^{3}}{\kappa^{3}}\right]^{-\frac{1}{2}}\,(S(p)_{\mu}S(p)^{\mu}-m^{2})b_{\mathbf{p^{*}}}^{\dag}\,e^{i(S(\omega_{p}^{*})t-S(\mathbf{p}^{*})\mathbf{x})}\label{EOMbdag}
\end{align}
\begin{align}
(\partial_{\mu}^{\dag}(\partial^{\mu})^{\dag}-m^{2})\phi^{\dag}= & \int\frac{d^{3}p}{\sqrt{2\omega_{p}}}\left[1+\frac{|p_{+}|^{3}}{\kappa^{3}}\right]^{-\frac{1}{2}}\,(S(S(p))_{\mu}S(S(p))^{\mu}-m^{2})a_{\mathbf{p}}^{\dag}\,e^{-i(S(\omega_{p})t-S(\mathbf{p})\mathbf{x})}\label{EOMadag}\\
+ & \int\frac{d^{3}p^{*}}{\sqrt{2|\omega_{p}^{*}|}}\left[1+\frac{|p_{+}|^{3}}{\kappa^{3}}\right]^{-\frac{1}{2}}\,(S(p^{*})_{\mu}S(p^{*})^{\mu}-m^{2})b_{\mathbf{p^{*}}}\,e^{i(\omega_{p}^{*}t-\mathbf{p}^{*}\mathbf{x})}.\label{EOMb}
\end{align}
Notice that \eqref{EOMa} is equivalent to \eqref{EOMadag} because
one can show that $S(S(p))_{\mu}S(S(p))^{\mu}=p_{\mu}p^{\mu}$, and
analogously \eqref{EOMb} is equivalent to \eqref{EOMbdag} because
$S(p^{*})_{\mu}=-S(p)_{\mu}$.

We find that with the
definition of the fields \eqref{fa} and \eqref{fb} the particle,
characterized by creation (annihilation) operator $a_{\mathbf{p}}$
($a_{\mathbf{p}}^{\dag}$) has the mass shell condition $p^{2}-m^{2}=0$,
while the antiparticle characterized by creation (annihilation) operator
$b_{\mathbf{p}}$ ($b_{\mathbf{p}}^{\dag}$) follows the mass-shell
condition $S(p)^{2}-m^{2}=0$. These mass-shells are identical, so that both the particle and the antiparticle have
the same rest mass, and the mass-shell manifold is in both cases the
same, but when we apply a Lorentz boost to a particle and an antiparticle
at rest with the same boost parameter, they would end up carrying
different momenta and energies. This leads to subtle deformation of
CPT symmetry, discussed in \cite{Arzano:2019toz} and \cite{Arzano:2020rzu}.

\section{Symmetries of the action}

 Let us now check that the above-defined fields transform properly under Poincar\'e and discrete symmetries, rendering the action \eqref{4.9} invariant.

\subsection{Poincar\'e symmetry of the action}

 In order to check the Poincar\'e invariance of the complex scalar field action\footnote{The paper \cite{Mercati:2011pv} provides a general abstract proof of Poincar\'e invariance of the $\kappa$-deformed complex scalar field action in 2 spacetime dimensions; here we show explicitly that the same holds in the particular of the theory considered here, in 4 dimensions.} \eqref{4.9} it is convenient to rewrite it in the momentum space where such invariance can be easily checked. As for the space-time action the procedure is much more involved and it is reported in Appendix \ref{Poincaresymm}.

Let us note that in order to turn the space-time action \eqref{4.9} to momentum space one we cannot use the on-shell field decomposition \eqref{fa}, \eqref{fb}, because the resulting momentum space action would contain the mass shell conditions as coefficients, which will make the action identically equal to zero. Therefore we use as a starting point the off-shell field decomposition
\begin{align}\label{fa0}
  \phi^{off}(x) &= \int_{\mathfrak{J}^+} \frac{d^4p}{p_4/\kappa}\, \xi(p)\, a_{p}\, e^{-i(p_0t-\mathbf{p}\mathbf{x})}  + \int_{\mathfrak{J}^-} \frac{d^4p^*}{|p^*_4|/\kappa} \, \xi(p^*)\, b^\dag_{p^*}\, e^{i(S(p^*_0)t-S(\mathbf{p}^*)\mathbf{x})}
\end{align}
where we include the additional factor (\ref{xi}) to make the momentum space action as simple as possible. In \eqref{fa0} we used the left-invariant measure (\ref{measure4d}) on the group manifold $\sf{AN}(3)$, and we restricted the range of integration in the first term to the positive energy $p_0>0$ subspace ${\mathfrak{J}^+}$
and to the energy $p^*_0<0$ subspace ${\mathfrak{J}^-}$ in the second term. This arrangement is analogous to the introduction of the $\theta$s in (\ref{phiPlusMinus}), but without the mass-shell restriction. The decomposition \eqref{fa0} can be further simplified observing that since $p^*$ is a dummy variable we can instead use the variables $p=-p^*$ in the second integral, so that we have
\begin{align}\label{fa1}
  \phi^{off}(x) &= \int_{\mathfrak{J}^+} \frac{d^4p}{p_4/\kappa}\, \left(1+\left(\frac{|p_+|}\kappa\right)^3\right)^{-1/2}\left( a_{p}\, e^{-i(p_0t-\mathbf{p}\mathbf{x})}  +  b^\dag_{-p}\, e^{-i(S(p_0)t-S(\mathbf{p})\mathbf{x})} \right)
\end{align}

The adjoint field has the form
\begin{align}\label{fb1}
  (\phi^{off})^\dag(x) &= \int_{\mathfrak{J}^+} \frac{d^4p}{p_4/\kappa}\, \left(1+\left(\frac{|p_+|}\kappa\right)^3\right)^{-1/2}\left( a^\dag_{p}\, e^{-i(S(p_0)t-S(\mathbf{p})\mathbf{x})} + b_{-p}\, e^{-i(p_0t-\mathbf{p}\mathbf{x})}\right)
\end{align}
Plugging these expressions to the action integral \eqref{4.9} after tedious computations, adjusting the free functions
we obtain the momentum space action in the form
\begin{equation}\label{momact}
  S =\frac12\,
	\int_{\mathfrak{J}^+}
	\frac{d^4p}{p_4/\kappa}\,(p_\mu p^\mu -m^2)
	a_{\mathbf{p}}^\dag a_{\mathbf{p}}
	+
	(S(p)_\mu S(p)^\mu -m^2)
	b_{\mathbf{p}} b_{\mathbf{p}}^\dag
\end{equation}%
It is clear from the action in the form \eqref{momact} above that the mass-shell of the `particle' is $p^2 =m^2$, while for the `antiparticle' it has the form $S(p)^2=m^2$, as discussed above.

Moreover it is straightforward to check its Poincar\'e invariance. The translations act on $a_{p}$, $b_{p}$ as phases; for the translation parameter $\bm\varepsilon$ we have
\begin{equation}\label{Ptrans}
  a_{p}\mapsto e^{i\varepsilon {p}}\,a_{p}\,,\quad b_{p}\mapsto e^{i\varepsilon {p}}\,b_{p}
\end{equation}
Next the action is clearly rotational invariant, if we assume that $a_p$, $b_{p}$ are scalar functions of the spacial momenta $\mathbf{p}$. It therefore remains to check the Lorentz invariance of the action. But since the action \eqref{momact} has the form of the standard undeformed momentum space action, the transformation properties of the creation and annihilation `operators' are just the standard ones $a_{p} \mapsto U(\Lambda) a_{p}U^{-1}(\Lambda) = a_{{\Lambda p}}$, where  $\Lambda p$ is the  Lorentz transformed four vector $ p$. Indeed
\begin{align}
U(\Lambda)	S U^{-1}(\Lambda)
	&=
	\frac{1}{2}
	\int_{\mathfrak{J}^+}
	d^4p\,
	(p_\mu p^\mu -m^2)
	U(\Lambda)a_{{p}}^\dag a_{{p}}U^{-1}(\Lambda)\nonumber\\
	&+
\frac{1}{2}
	\int_{\mathfrak{J}^+}
	d^4p\,	(S(p)_\mu S(p)^\mu -m^2)
U(\Lambda)	b_{{p}}^\dag b_{{p}}U^{-1}(\Lambda)\nonumber\\
&=
\frac{1}{2}
	\int_{\mathfrak{J}^+}
	d^4(\Lambda p)\,
	((\Lambda p)_\mu (\Lambda p)^\mu -m^2)
	a_{\Lambda {p}}^\dag a_{\Lambda {p}}\nonumber\\
	&+
\frac{1}{2}
	\int_{\mathfrak{J}^+}
	d^4(\Lambda p)\,	(S(\Lambda p)_\mu S(\Lambda p)^\mu -m^2)
b_{\Lambda {p}}^\dag b_{\Lambda {p}}\nonumber\\ &= S\nonumber
\end{align}
This completes the proof of Poincar\'e invariance of the action \eqref{momact}.

\section{Discrete symmetries}\label{CPT}

There are three discrete symmetries: parity $\cal P$, time reversal $\cal T$, and charge conjugation $\cal C$. In each case we will first shortly recall their action on the undeformed field  with decomposition
\begin{equation}\label{p1}
 \phi(t,\mathbf{x}) = \int \frac{d^3p}{\sqrt{2\omega_p}}\, a_\mathbf{p}\, e^{-i(\omega_pt - i{\mathbf{p}}{\mathbf{x}})}+ b^\dag_\mathbf{p}\, e^{i(\omega_p t -{\mathbf{p}}{\mathbf{x}})}
\end{equation}
and then generalize it to the case of the deformed fields \eqref{fa} and \eqref{fb}. For parity and time reversal we have spacetime concepts to guide us, and therefore we consider these two first.

\subsection{ Parity}

The parity operator $\cal P$ acts on space coordinates as an inversion $x=(t,\mathbf{x})\rightarrow x' = (t,-\mathbf{x})$. For the complex scalar quantum field, we define the parity operator as
\begin{equation}\label{p2}
{\cal P}  \phi(t,\mathbf{x}){\cal P}^{-1} =\int \frac{d^3p}{\sqrt{2\omega_p}}\,{\cal P} a_\mathbf{p}{\cal P}^{-1} \, e^{-i(\omega_pt - {\mathbf{p}}{\mathbf{x}})}+{\cal P} b^\dag_\mathbf{p}{\cal P}^{-1} \, e^{i(\omega_pt -{\mathbf{p}}{\mathbf{x}})}\equiv\phi(t,-\mathbf{x})
\end{equation}
and using \eqref{p1} we see that for the creation/annihilation operators\footnote{Here and below we ignore a possible phase factor that may be present in the definition.}
\begin{equation}\label{p3}
  {\cal P} a_{\mathbf{p}}{\cal P}^{-1} = a_{-\mathbf{p}}\,,\quad {\cal P} b_{\mathbf{p}}{\cal P}^{-1} = b_{-\mathbf{p}}
\end{equation}

Turning to the deformed case we notice first that the spacetime transformation $\hat{\mathbf{x}}\mapsto -\hat{\mathbf{x}} $ leaves the defining commutator \eqref{0.1} invariant and therefore is compatible with the form of $\kappa$-Minkowski non-commutativity. Further, the positive and negative energy fields $\phi_{(\pm)}(x)$ can be considered separately. For the positive energy part we can use exactly the same considerations as in the case of the undeformed field above. Since
$$
S(p_0, -\mathbf{p})_0=S(p_0, \mathbf{p})_0 , \quad S(p_0, -\mathbf{p})_i= -S(p_0, \mathbf{p})_i
$$
this is also true for the negative energy fields and thus
 we can readily define
\begin{equation}\label{p4}
  {\cal P} a_{\mathbf{p}}{\cal P}^{-1} = a_{-\mathbf{p}}\,,\quad {\cal P} b_{\mathbf{p^*}}{\cal P}^{-1} = b_{-\mathbf{p^*}}
\end{equation}
and
\begin{equation}\label{p5}
  {\cal P} a^\dag_{\mathbf{p}}{\cal P}^{-1} = a^\dag_{-\mathbf{p}}\,,\quad {\cal P} b^\dag_{\mathbf{p^*}}{\cal P}^{-1} = b^\dag_{-\mathbf{p^*}}
\end{equation}

\subsection{ Time reversal}

Next we consider the time reversal $\cal T$, which changes the time direction $x=(t,\mathbf{x})\rightarrow x' = (-t,\mathbf{x})$ and
 \begin{equation}\label{t1}
{\cal T}  \phi(t,\mathbf{x}){\cal T}^{-1} =\phi(-t,\mathbf{x})
\end{equation}
 It should be remembered that the operator $\cal T$ is anti-hermitian ${\cal T}i {\cal T}^{-1} = -i$, and  we have
\begin{align}
{\cal T}  \phi(t,\mathbf{x}){\cal T}^{-1} &=\int \frac{d^3p}{\sqrt{2\omega_p}}\,{\cal T} a_\mathbf{p}\, e^{-i(\omega_pt - i{\mathbf{p}}{\mathbf{x}})}{\cal T}^{-1}+{\cal T} b^\dag_\mathbf{p}\, e^{i(\omega_pt -{\mathbf{p}}{\mathbf{x}})}{\cal T}^{-1}\nonumber\\
&= \int \frac{d^3p}{\sqrt{2\omega_p}}\,{\cal T} a_\mathbf{p}{\cal T}^{-1}\, e^{i(\omega_pt - i{\mathbf{p}}{\mathbf{x}})}+{\cal T} b^\dag_\mathbf{p}{\cal T}^{-1}\, e^{-i(\omega_pt -{\mathbf{p}}{\mathbf{x}})} = \phi(-t,\mathbf{x})\label{t2}
\end{align}
We find that
\begin{equation}\label{t3}
  {\cal T} a_{\mathbf{p}}{\cal T}^{-1} = a_{-\mathbf{p}}\,,\quad {\cal T} b_{\mathbf{p}}{\cal T}^{-1} = b_{-\mathbf{p}}.
\end{equation}

Let us now discuss the deformed case. We start noticing that as a consequence of anti-hermiticity of $\cal T$ the defining algebra \eqref{0.1} is again invariant, so that we see that $\kappa$-Minkowski space is both parity and time reversal invariant. Turning to fields we again see that the classical reasoning can be verbatim repeated in the case of time reversal as well and we end up with
\begin{equation}\label{t4}
  {\cal T} a_{\mathbf{p}}{\cal T}^{-1} = a_{-\mathbf{p}}\,,\quad {\cal T} b_{\mathbf{p^*}}{\cal T}^{-1} = b_{-\mathbf{p^*}}
\end{equation}
and
\begin{equation}\label{t5}
  {\cal T} a^\dag_{\mathbf{p}}{\cal T}^{-1} = a^\dag_{-\mathbf{p}}\,,\quad {\cal T} b^\dag_{\mathbf{p^*}}{\cal T}^{-1} = b^\dag_{-\mathbf{p^*}}
\end{equation}

\subsection{ Charge conjugation}

The symmetry that exchanges particles with antiparticles does not have any spacetime counterparts and since it changes the charge it is called charge conjugation. The charge conjugation operator ${\cal C}$ acting on the field produces its conjugation
\begin{equation}\label{c1}
{\cal C}  \phi(t,\mathbf{x}){\cal C}^{-1} =\phi^\dag(t,\mathbf{x})
\end{equation}
therefore
\begin{equation}\label{c2}
  {\cal C}  \phi(t,\mathbf{x}){\cal C}^{-1} = \int \frac{d^3p}{\sqrt{2\omega_p}}\,{\cal C} a_\mathbf{p}{\cal C}^{-1}\, e^{-i(\omega_pt - i{\mathbf{p}}{\mathbf{x}})}+{\cal C} b^\dag_\mathbf{p}{\cal C}^{-1}\, e^{i(\omega_pt -{\mathbf{p}}{\mathbf{x}})}  =\phi^\dag(t,\mathbf{x})
\end{equation}
and we have
\begin{equation}\label{c3}
  {\cal C} a_{\mathbf{p}}{\cal C}^{-1} = b_{\mathbf{p}}
\end{equation}

Let us now consider the deformed field. Take the $\phi_{(+)}$ component first
\begin{equation}\label{c4}
  {\cal C}  \phi_{(+)}(t,\mathbf{x}){\cal C}^{-1} = \int \frac{d^3p}{\sqrt{2\omega_p}} \left[1+\frac{|p_+|^3}{\kappa^3}\right]^{-\frac{1}{2}}\,{\cal C} a_\mathbf{p}{\cal C}^{-1}\, e^{-i(\omega_pt - i{\mathbf{p}}{\mathbf{x}})}
\end{equation}
On the other hand we have
\begin{equation}\label{c5}
 \phi^\dag_{(-)}(x)= \int \frac{d^3p^*}{\sqrt{2|\omega^*_p|}} \left[1+\frac{|p_+|^3}{\kappa^3}\right]^{-\frac{1}{2}} \, b_\mathbf{p^*}\, e^{i(\omega^*_pt-\mathbf{p}^*\mathbf{x})}
\end{equation}
so that we can conclude that
\begin{equation}\label{c6}
 {\cal C} a_\mathbf{p}{\cal C}^{-1} =  b_{\mathbf{p}^*}
\end{equation}
Analogously, for the $\phi_{(-)}$ component we have
\begin{equation}\label{c7}
{\cal C}   \phi_{(-)}(x) {\cal C}^{-1} =\int \frac{d^3p^*}{\sqrt{2|\omega^*_p|}} \left[1+\frac{|p_+|^3}{\kappa^3}\right]^{-\frac{1}{2}} \,{\cal C} b^\dag_\mathbf{p^*}{\cal C}^{-1}\, e^{i(S(\omega^*_p)t-S(\mathbf{p}^*)\mathbf{x})}
\end{equation}
and
\begin{equation}\label{c8}
  \phi_{(+)}^\dag(x) = \int \frac{d^3p}{\sqrt{2\omega_p}}\,\left[1+\frac{|p_+|^3}{\kappa^3}\right]^{-\frac{1}{2}} a^\dag_\mathbf{p}\, e^{-i(S(\omega_p)t-S(\mathbf{p})\mathbf{x})}
\end{equation}
So that
\begin{equation}\label{c9}
  {\cal C} b^\dag_{\mathbf{p}^*}{\cal C}^{-1}=a^\dag_\mathbf{p}
\end{equation}

It should be stressed that this simple transformation rules of the field $\phi$ with respect to charge conjugation is a result of the use of the second (starred) copy of momentum space and of the particular arrangement of the components $\phi_{(\pm)}(x)$ and $\phi^\dag_{(\pm)}(x)$. In particular, the field constructed in \cite{Freidel:2007hk} and many other papers on this topic does not transform nicely under charge conjugation. It should be added also that the deformed action of discrete symmetries $\cal P$, $\cal T$, and $\cal C$ leads to the form of the $\cal CPT$ operator $\Theta$ anticipated in the paper \cite{Arzano:2019toz}, although the action of parity and time reversal differ from that proposed in \cite{Arzano:2016egk}.

\section{Conserved charges and symplectic structure}\label{Sect.charges}

 In this section we derive the conserved charges and symplectic structure associated with our free complex scalar field theory defined by the action
 \begin{align}\label{VII.1}
 S= \frac{1}{2}(S_1 + S_2) &=  \frac12\,	\int_{\mathbb{R}^4} d^4x \,\, (\partial^\mu \phi)^\dag \star (\partial_\mu \phi) - m^2 \phi^\dag \star \phi\nonumber\\
 &+ \frac12\,\int_{\mathbb{R}^4} d^4x \,\ (\partial_\mu \phi) \star (\partial^\mu \phi)^\dag  - m^2 \phi \star \phi^\dag
 \end{align}
 Both are given in terms of the appropriate boundary integrals, and reflect, respectively, the symmetries of the theory (charges) and its kinematics (symplectic structure). Our starting point here will be the variations of the actions computed above, eqs.\ \eqref{4.35}--\eqref{4.44}. Assuming field equations in the bulk these variations are just the boundary terms, which become conserved charges in the case of field variations corresponding to symmetries of the action and Liouville form, for generic variations.

\subsection{Conserved charges}

On-shell the variation of the action reduces to the boundary term and we define the conserved charges associated with the field transformation that leaves the action invariant $\delta_S\phi$, $\delta_S \phi^\dag$ as usual as an integral over the constant time surface
\begin{equation}\label{VII.6}
  {\cal P}_S =\frac12\, \int d^3x\, \Pi_1^0 \star \delta_S \phi +\delta_S\phi^\dag \star \left(\Pi_1^0\right)^\dag +  \delta_S \phi \star \Pi_2^0+ \left(\Pi_2^0\right)^\dag	\star	\delta_S\phi^\dag
\end{equation}
In the case of translational symmetry, for which
\begin{equation}\label{VII.7}
  \delta_S\phi = d\phi = \epsilon^A\partial_A\phi
\end{equation}
we find
\begin{equation}\label{VII.8}
  {\cal P}_A =\frac12\, \int d^3x\, \tensor{T}{_1^0_{A}} + \tensor{T}{_2^0_{A}}
\end{equation}
where the relevant components of the energy-momentum tensor are
\begin{equation}\label{VII.9}
  \tensor{T}{_1^0_{A}}
	=
	-
	\partial_A \Pi_1^0 \star \phi
	+
	\partial_A\phi^\dagger \star {\Pi_1^\dagger}^0
\end{equation}
and
\begin{equation}\label{VII.9a}
  \tensor{T}{_2^0_{A}}
	=
	-
	\phi \star \partial_A \Pi_2^0
	+
	{\Pi_2^\dagger}^0 \star \partial_A\phi^\dagger
\end{equation}

Now we use the field decomposition \eqref{fa}, \eqref{fb} to find the expression for conserved translational charges ${\cal P}_A$ \eqref{VII.8} in momentum space. After tedious computation one finds that the time dependent terms cancel as they should and the conserved charges have the form
\begin{align}\label{VII.10}
	\mathcal{P}_0
	=
	- \frac{1}{2} \int
	d^3p\,
	a_\mathbf{p}^\dag\,
	a_\mathbf{p}\,
	S(\omega_p)
	\left[
	1 - \frac{\xi(p)^2 \mathbf{p}^2}{\omega_p p_+}
	\right]
	\frac{p_4}{\kappa}
	-
	b_{\mathbf{p}^*}\,
	b_{\mathbf{p}^*}^\dag\,
	\omega_p
	\left[
	1 - \frac{\xi(p)^2 \mathbf{p}^2}{\omega_p p_+}
	\right]
	\frac{p_4}{\kappa}
\end{align}
\begin{align}\label{VII.11}
	\mathcal{P}_i
	=
	\frac{1}{2}
	\int d^3p \,
	a_\mathbf{p}^\dag\,
	a_\mathbf{p}\,
	S(\mathbf{p})_i
	\left[
	1 - \frac{\xi(p)^2 \mathbf{p}^2}{\omega_p p_+}
	\right]
	\frac{p_4}{\kappa}
	-
	b_{\mathbf{p}^*}\,
	b_{\mathbf{p}^*}^\dag\,
	\mathbf{p}_i
	\left[
	1 - \frac{\xi(p)^2 \mathbf{p}^2}{\omega_p p_+}
	\right] \frac{p_4}{\kappa}
\end{align}
\begin{align}\label{VII.12}
	\mathcal{P}_4
	=
	-\frac{1}{2}
	\int d^3p
	\left(p_4 - \kappa\right)
	\left\{
	a_\mathbf{p}^\dag\,
	a_\mathbf{p}\,
	\left[
	1 - \frac{\xi(p)^2 \mathbf{p}^2}{\omega_p p_+}
	\right]
	\frac{p_4}{\kappa}
	-
	b_{\mathbf{p}^*}\,
	b_{\mathbf{p}^*}^\dag\,
	\left[
	1 - \frac{\xi(p)^2 \mathbf{p}^2}{\omega_p p_+}
	\right]
	\frac{p_4}{\kappa}
	\right\}.
\end{align}

\subsection{Symplectic structure}

To compute the symplectic structure of our theory we use the covariant phase space approach \cite{Crnkovic:1987tz}, \cite{Lee:1990nz}, \cite{Iyer:1994ys}, which makes it possible to to straightforwardly derive it from the action preserving all the relavant symmetries. To compute the symplectic structure we must return to \eqref{4.36} and \eqref{4.41}. Defining the Liouville form $\theta$ as a boundary term in the variation of the action on-shell, for generic variation of the field $\delta\phi$, $\delta\phi^\dag$ we find
\begin{equation}\label{VII.13}
  \theta = \theta_1 +\theta_2 =-\frac12\, \int_{\mathbb{R}^3} d^3x
	\left(
	\Pi^0_1 \star \delta\phi
	+
	\delta\phi^\dag \star \left(\Pi^0_1\right)^\dag
	+
	\delta\phi \star \Pi^0_2
	+
	\left(\Pi^0_2\right)^\dag \star \delta\phi^\dag
	\right)
\end{equation}
To find the symplectic form, which will lead to the Poisson bracket of field coefficients $a$ and $b$ and in turn to the creation/annihilation operators commutators, we have to compute $\delta\theta$ and express the result using the momentum space decomposition \eqref{fa}, \eqref{fb}. We find
\begin{align}\label{VII.14}
	\delta\theta_1 + \delta\theta_2
	&=
	-\frac{i}{2}\int d^3p \,
	a_\mathbf{p}\wedge
	a_\mathbf{p}^\dag\,
	\left[
	1 - \frac{\xi(p)^2 \mathbf{p}^2}{\omega_p p_+}
	\right]
	\frac{p_4}{\kappa}
	-
	\xi(p)^2
	b_{\mathbf{p}^*}^\dag\wedge
	b_{\mathbf{p}^*}\,
	\left[
	1 - \frac{\xi(p)^2 \mathbf{p}^2}{\omega_p p_+}
	\right]
	\frac{p_4}{\kappa}.
\end{align}
which implies the following Poisson brackets
\begin{align}
	\left\{
	a_\mathbf{p}, a_\mathbf{q}^\dag
	\right\}
	&=i
	\frac{\kappa}{p_4}
	\frac{2}{1 - \frac{\xi(p)^2 \mathbf{p}^2}{\omega_p p_+}}
	\delta(\mathbf{p}-\mathbf{q}) \label{VII.15}\\
	\left\{
	b_{\mathbf{p}^*}, b_{\mathbf{q}^*}^\dag
	\right\}
	&=i
	\frac{\kappa}{p_4}
	\frac{2}{1 - \frac{\xi(p)^2 \mathbf{p}^2}{\omega_p p_+}}
	\delta(\mathbf{p}-\mathbf{q}) . \label{VII.16}
\end{align}

\section{Towards quantum theory}

In this section we will construct the one particle states in quantum field theory. At this stage we cannot go any further, in particular we cannot construct many-particles states and investigate their properties, because this would require knowing details of the coproduct properties of creation and annihilation operators, i.e. how they act on tensor product of states.

In quantum theory the Poisson brackets \eqref{VII.15}, \eqref{VII.16} become commutators (from now on we stop distinguishing $\mathbf{p}$ from $\mathbf{p}^*$)
\begin{align}
	\left[
	a_\mathbf{p}, a_\mathbf{q}^\dag
	\right]
	&= \frac{\kappa}{p_4}
	\frac{2}{1 - \frac{\xi(p)^2 \mathbf{p}^2}{\omega_p p_+}}
	\delta(\mathbf{p}-\mathbf{q}) \label{VIII.1}\\
	\left[
	b_{\mathbf{p}}, b_{\mathbf{q}}^\dag
	\right]
	&= \frac{\kappa}{p_4}
	\frac{2}{1 - \frac{\xi(p)^2 \mathbf{p}^2}{\omega_p p_+}}
	\delta(\mathbf{p}-\mathbf{q}). \label{VIII.2}
\end{align}

We define the vacuum $\left|0\right>$ that satisfies the condition
\begin{equation}\label{VIII.2a}
a_\mathbf{p}\left|0\right> = b_\mathbf{p}\left|0\right> =0
\end{equation}
Then we define the one-particle and one-antiparticle states
\begin{align}
  \left|\mathbf{p}\right>_a &\equiv a^\dag_\mathbf{p}\left|0\right> \label{VIII.3}\\
  \left|\mathbf{p}\right>_b &\equiv b^\dag_\mathbf{p}\left|0\right> \label{VIII.4}
\end{align}
Now we are ready to present the most important result of this investigations. Consider the state $\left|\mathbf{p}\right>_a$, \eqref{VIII.3}. Its momentum can be computed by acting with the momentum operator $\mathcal{P}_i$, \eqref{VII.11} on it. Using the commutational relation \eqref{VIII.1} we find
\begin{equation}\label{VIII.5}
 \mathcal{P}_i \left|\mathbf{p}\right>_a = -S(p)_i\, \left|\mathbf{p}\right>_a
\end{equation}
Analogously, for the one-antiparticle state $\left|\mathbf{p}\right>_b$, \eqref{VIII.3}, using the commutational \eqref{VIII.2} we get
\begin{equation}\label{VIII.6}
 \mathcal{P}_i \left|\mathbf{p}\right>_b = p_i\, \left|\mathbf{p}\right>_b
\end{equation}
In exactly the same manner we can use the Hamiltonian \eqref{VII.10} to compute the energy of the one particle states, obtaining
\begin{equation}\label{VIII.5a}
 \mathcal{P}_0 \left|\mathbf{p}\right>_a = -S(\omega_p)\, \left|\mathbf{p}\right>_a
\end{equation}
and
\begin{equation}\label{VIII.6a}
 \mathcal{P}_0 \left|\mathbf{p}\right>_b = \omega_p\, \left|\mathbf{p}\right>_b
\end{equation}
Therefore, one-particle and one-antiparticle states belong to the same mass-shell manifold, since
\begin{equation}\label{VIII.7}
  \omega_p^2 - \mathbf{p}^2=m^2 = S(\omega_p)^2 - S(\mathbf{p})^2
\end{equation}
but  $\mathbf{p}$ and $S(\mathbf{p})$ are, in general different points on this manifold, with a single exception being the case $\mathbf{p}=S(\mathbf{p})=0$, $\omega_p = -S(\omega_p)=m$.

Finally, the momentum ${\cal P}_4$ measures, essentially, the deformed charge of the state
\begin{equation}\label{VIII.5b}
 \mathcal{P}_4 \left|\mathbf{p}\right>_a = (\sqrt{\kappa^2 + m^2}-\kappa)\, \left|\mathbf{p}\right>_a
\end{equation}
and
\begin{equation}\label{VIII.6b}
 \mathcal{P}_4 \left|\mathbf{p}\right>_b = -(\sqrt{\kappa^2 + m^2}-\kappa)\, \left|\mathbf{p}\right>_b
\end{equation}

Therefore the one-particle state carries the momentum $-S(p)_i$, while the one-antiparticle state has the momentum $p_i$. But according to \eqref{c9} the latter is the ${\cal C}$ (and also ${\cal CPT}$) of the former
\begin{equation}\label{VIII.7a}
{\cal C}  \left|\mathbf{p}\right>_b ={\cal C} b^\dag_\mathbf{p}{\cal C}^{-1}\, {\cal C} \left|0\right> ={\cal C} b^\dag_\mathbf{p}{\cal C}^{-1} \left|0\right> =a^\dag_\mathbf{p}\left|0\right> =\left|\mathbf{p}\right>_a
\end{equation}
Therefore, as anticipated in  Section \ref{CSF} the charge conjugation (and  ${\cal CPT}$) transforms a particle into an antiparticle with different momentum. This transformation has the remarkable property that the rest mass of the particle and antiparticle is the same. The phenomenological consequences of this have been recently discussed in \cite{Arzano:2019toz}, \cite{Arzano:2020rzu}.

\section{Summary and conclusions}
We laid down the basic ingredients for the construction of a complex field theory on $\kappa$-Minkowski space covariant under the action of deformed relativistic symmetries described by the $\kappa$-Poincar\'e algebra. The guiding principle which we followed in the definition of the field and its action was the requirement of an appropriate transformation of the former under the action of discrete symmetries. The main upshot of our construction is that the four-momenta of particles and anti-particles states related by charge conjugation $\cal C$ are not identical, and given by eq.\ \eqref{VIII.5}--\eqref{VIII.6b}. After deriving the equations of motions from the deformed action we worked the action of Poincar\'e symmetries on the field both from a coordinate and momentum space perspective and then moved onto the description of the action of discrete symmetries. The last part of our work was devoted to the analysis of the symplectic structure of the theory which allowed us to derive the conserved charges associated to the deformed translation symmetries. This also made possible to write down the Poisson brackets of the expansion coefficients of the field which upon quantization become creation and annihilation operators. With these we were able to characterize the energy and momentum of one-particle and anti-particle states and write down the action of discrete symmetries on them which showed that the CPT operator maps particle states into anti-particle states with a different momentum. This important result could have non-trivial phenomenological consequences which might be relevant for experimental searches of Planck scale effects \cite{Arzano:2019toz, Arzano:2020rzu}.

There are several open issues that we are going to address in the future publications. First it does seem that the particle state and its associated charge conjugated anitiparticle one have different momenta, and it is not trivial to define the real scalar field. We will return to it in the forthcoming publications. The main open issue at the quantum level concerns the construction of a Fock space on which the commutators that we derived for creation and annihilation operators can act, mapping multiparticle states given by appropriately symmetrized tensor products of one-particle states consistent with the non-trivial co-product and covariant under the action of the $\kappa$-Poincar\'e algebra. This is notoriously a thorny issue which has not yet found a satisfactory answer \cite{Arzano:2007ef,Young:2007ag,Govindarajan:2008qa,Young:2008zg,Arzano:2008bt,Daszkiewicz:2008bm} and which we hope we will be able to successfully address within the approach to field theory proposed in this work. The satisfactory solution of this problem is the major prerequisite for the construction of the interacting $\kappa$-deformed quantum field theory and $\kappa$-deformed standard model, which is our ultimate goal in the research project that the present paper is the first step of.

\begin{acknowledgments}
For JKG, GR, and JU  this work was supported by funds provided by the National Science Center, project number  2019/33/B/ST2/00050 and for JKG and JU  also by the project number 2017/27/B/ST2/01902.
\end{acknowledgments}

\appendix

\section{Lorentz transformations of antipode}\label{AppS}

The antipodes were defined in \eqref{0.4} and are given by the following expressions
\begin{equation}\label{aa1}
  S(p_0) = -p_0 + \frac{\mathbf{p}^2}{p_0+p_4} = \frac{\kappa^2}{p_0+p_4}-p_4\,,\quad S(\mathbf{p}) =-\frac{\kappa \mathbf{p} }{p_0+p_4}\,,\quad S(p_4) = p_4
\end{equation}
The action of Lorentz boost transformation on the antipode is defined as \eqref{0.4b}
\begin{equation}\label{aa2}
  L \triangleright S(p) \equiv S(L\triangleright p)\,,\quad p_0>0
\end{equation}
Let us investigate properties of this transformation in the case of a infinitesimal Lorentz transformation with parameter $\xi^i$
\begin{equation}\label{aa3}
  \delta_\xi p_i = \xi_i\, p_0\,,\quad \delta_\xi p_0 = \xi^i\, p_i
\end{equation}
Remembering that $p_4$ is Lorentz-invariant using \eqref{aa2} we find
\begin{equation}\label{aa4}
  \delta_\xi S(p_0) = -\xi^i\, p_i +\frac{2\xi^i\, p_i\, p_0 }{p_0+p_4} -\frac{\mathbf{p}^2}{(p_0+p_4)^2}\, \xi^i\, p_i = \zeta^i\, S(p_i)
\end{equation}
where we introduce a momentum-dependent infinitesimal parameter
\begin{equation}\label{aa5}
  \zeta^i = \xi^i \frac{\kappa }{p_0+p_4}
\end{equation}
Thus the Lorentz transformation of the zero component of the antipode is an ordinary Lorentz transformation, with parameter $\zeta^i$.

For the spacial component we have a more complicated expression.
\begin{align}
  \delta_\xi S(p_i) &= -\frac{\kappa \xi_i\, p_0 }{p_0+p_4} + \frac{\kappa {p}_i }{(p_0+p_4)^2}\, \xi^j\, p_j\nonumber\\
  &= \zeta_i S(p_0) +\frac{\kappa }{(p_0+p_4)^2}\left(p_i\xi^jp_j - \xi_i \mathbf{p}^2\right)\label{aa6}
\end{align}
The first term here is again the standard Lorentz transformation with parameter $\zeta^i$. The second term is an infinitesimal rotation of $S(p_i)$ with the parameter
$$
\rho^j =\epsilon^j{}_{kl}\,\xi^k\, p^l
$$
so that, finally
\begin{equation}\label{aa7}
  \delta_\xi S(p_i) = \zeta_i S(p_0) + \epsilon_i{}^{jk}\,\rho_j \, S(p_k)
\end{equation}
Since under Lorentz boost transformation of momenta the components of the antipode transform  under a combination of boost and rotation it is clear that the components of the antipode satisfy the same mass shell condition as the components of the original momenta.

\section{Integration by parts}\label{Intbyparts}

In this Appendix we derive the $\star$-integration by parts formula, which is necessary do derive field equations from the action \eqref{4.9}.

The starting point is provided by the coproduct rules for the $\kappa$-Poincar\'e algebra in the classical basis $(p_0,p_i,p_4)$ \cite{Freidel:2007hk}, \cite{Kowalski-Glikman:2017ifs}
\begin{align}
\Delta p_i &= \frac{1}{\kappa} p_i \otimes (p_0+p_4)
+ 1 \otimes p_i \label{4.10}\\
\Delta p_0 = \frac{1}{\kappa} p_0 \otimes (p_0+p_4)
&+
\sum p_k (p_0+p_4)^{-1} \otimes p_k
+
\kappa (p_0+p_4)^{-1} \otimes p_0 \label{4.11}\\
\Delta p_4 = \frac{1}{\kappa} p_4 \otimes (p_0+p_4)
&-
\sum p_k (p_0+p_4)^{-1} \otimes p_k
-
\kappa (p_0+p_4)^{-1} \otimes p_0. \label{4.12}
\end{align}
Notice that the coproduct relations are an immediate consequence of \eqref{0.15}.  The coproducts tell us how the momentum operators act on star products of two functions. Since momenta are spacetime derivatives $p_0 = i\partial_0$, $p_i = i\partial_i$ these equations tell us how derivatives act on the star products of functions on Minkowski space, defining in this way the modified Leibniz rules. In the calculation below we use the short-hand notation $p_+ \rightarrow \Delta_+=i\partial_0+p_4 = i\partial_0 + (\kappa + i\partial_4)$, where the nonlocal operator $p_4$ is expressed in terms of the corresponding derivatives as $p_4 = \sqrt{\kappa^2 -\partial_0^2 + \partial_i^2}$. Equations \eqref{4.10}, \eqref{4.11}, \eqref{4.12} then imply
\begin{align}
\partial_0 (\phi\star \psi) = \frac{1}{\kappa}
	(&\partial_0\phi) \star (\Delta_+\psi)
	+ \kappa
	(\Delta_+^{-1}\phi)\star(\partial_0\psi)
	+i
	(\Delta_+^{-1}\partial_i\phi)\star(\partial_i\psi) \label{4.13}\\
	&\partial_i (\phi\star \psi) =
	\frac{1}{\kappa}(\partial_i \phi)\star (\Delta_+\psi)
	+
	\phi\star(\partial_i\psi) \label{4.14}\\
	& \qquad
	\Delta_+(\phi \star \psi) = \frac{1}{\kappa}(\Delta_+\phi) \star (\Delta_+\psi). \label{4.15}
\end{align}
Furthermore defining the adjoint derivative
\begin{equation}\label{4.16a}
  (\partial_A\phi)^\dag \equiv \partial_A^\dag \phi^\dag\,,\quad A=(\mu, 4, +)
\end{equation}
and using equation \eqref{0.4} we have
\begin{align}\label{4.16}
\partial_i^\dagger = \kappa \Delta_+^{-1} \partial_i,
\qquad
\partial_0^\dagger = \partial_0 - i\Delta_+^{-1}\bm{\partial}^2,
\qquad
\partial_4^\dagger = -\partial_4,
\qquad
\Delta_+^\dagger = \kappa^2 \Delta_+^{-1}.
\end{align}
We now use eq. \eqref{4.13}, \eqref{4.14}, \eqref{4.15}, \eqref{4.16} to obtain the expressions needed for the integration by parts of expressions of the form $(\partial_\mu\phi)^\dag \star\partial^\mu \psi$ and $(\partial_\mu \psi) \star (\partial^\mu\phi)^\dag$. With some algebra we find
\begin{align}
(\partial_i\phi)^\dag \star (\partial_i\psi)
	=
	\partial_i\left[
	(\partial_i\phi)^\dagger \star \psi
	\right]
	-
	\frac{\Delta_+}{\kappa}
	\left[(\bm{\partial}^2\phi)^\dag \star \psi \right].\label{4.25}
\end{align}
Similarly
\begin{align}
(\partial_0\phi)^\dag \star (\partial_0 \psi)
	=
	\frac{\partial_0}{\kappa}\left[
	(\Delta_+(\partial_0\phi)^\dag) \star \psi
	\right]
	-
	i\partial_i[(\Delta_+^{-1}\partial_i{\partial_0\phi})^\dag\star \psi]
	-
	\frac{\Delta_+}{\kappa}\left[
	(\partial_0^2\phi)^\dag \star \psi
	\right].\label{4.27}
\end{align}
Notice that using this convention, equations \eqref{4.25} and \eqref{4.27} are still fine substituting $\phi^\dag$ with any other quantity (because the above derivations do not use in any way the presence of the ${}^\dag$ over $\phi$), and therefore can be used regardless of the combination of fields to which they can be applied.

The hermitian conjugates of equations \eqref{4.25}, \eqref{4.27} take the form
\begin{align}
	&(\partial_i\psi)^\dag \star (\partial_i\phi)
	=
	\partial_i^\dag \left[
	\psi^\dag \star (\partial_i\phi)
	\right]
	-
	\frac{\kappa}{\Delta_+}
	\left[\psi^\dag \star (\bm{\partial}^2\phi) \right] \label{4.32} \\
	 (\partial_0\psi)^\dag \star (\partial_0 \phi&)
	 =
{\partial_0^\dag}\left[
	 \psi^\dag \star ({\kappa}\Delta_+^{-1}\,\partial_0\phi)
	 \right]
	 +
	 i\partial_i^\dag[\psi^\dag \star (\Delta_+^{-1}\partial_i{\partial_0\phi}) ]
	 -
	 \frac{\kappa}{\Delta_+}\left[
	 \psi \star (\partial_0^2\phi)
	 \right]. \label{4.33}
\end{align}
Finally, we will also need the following identity
\begin{align}
m^2 \phi^\dag \star \psi
&=
-\left(\frac{\Delta_+}{\kappa} - 1\right)(m^2 \phi^\dag \star \psi )
+
\frac{\Delta_+}{\kappa}
(m^2 \phi^\dag \star \psi)  \nn \\
&=
-
\frac{i\partial_0}{\kappa}
(m^2 \phi^\dag \star \psi)
-
\frac{i\partial_4}{\kappa}
(m^2 \phi^\dag \star \psi)
+
\frac{\Delta_+}{\kappa}
(m^2 \phi^\dag \star \psi)\label{4.30}
\end{align}
and its hermitian conjugate
\begin{align}
m^2 \psi^\dag \star \phi
	&=
	+
	\frac{i\partial_0^\dag }{\kappa}
	(m^2 \psi^\dag \star \phi)
	+
	\frac{i\partial_4^\dag }{\kappa}
	(m^2 \psi^\dag \star \phi)
	+
	\frac{\kappa}{\Delta_+}
	(m^2 \psi^\dag \star \phi). \label{4.34}
\end{align}
For the opposite ordering we have instead
\begin{align}\label{4.34-1}
	(\partial_i \psi) \star (\partial_i \phi)^\dag
	=
	\kappa \partial_i (\psi \star [\Delta_+^{-1}(\partial_i \phi)^\dag])
	-
	\psi \star (\bm{\partial}^2 \phi)^\dag.
\end{align}
\begin{align}\label{4.34-2}
	(\partial_0 \psi) \star (\partial_0 \phi)^\dag
	=&
	\partial_0 (\psi \star [\kappa\Delta_+^{-1}(\partial_0\phi)^\dag])
	-
	i
	\partial_i (\psi \star [\Delta_+^{-1}\partial_i(\partial_0 \phi)^\dag])
	-
	[\psi \star (\partial_0^2 \phi)^\dag] \nn \\
	&+
	\left(\frac{i}{\kappa}\partial_0 + i\frac{\partial_4}{\kappa}\right)
	[\psi \star (\partial_0^2 \phi)^\dag].
\end{align}

\section{Poincar\'e symmetry of the action -- spacetime approach}\label{Poincaresymm}

We want to discuss the invariance of the action~(\ref{4.9})
under $\kappa$-Poincar\'e transformations. As a first step, let
us notice that it is equivalent to
\begin{equation}
S=-\frac{1}{2}\int_{\mathbb{R}^{4}}d^{4}x\left[\phi^{\dagger}\star\partial_{\mu}\partial^{\mu}\phi+\phi\star\left(\partial_{\mu}\partial^{\mu}\phi\right)^{\dagger}+m^{2}\left(\phi^{\dagger}\star\phi+\phi\star\phi^{\dagger}\right)\right].\label{action}
\end{equation}
This is easy to see using~(\ref{4.34-1}) and~(\ref{4.34-2}) for integrating by parts the second term, and~(\ref{4.32}) and~(\ref{4.33}) for the first term, in the action~(\ref{4.9}),  since the Lagrangians are the same up to a total divergence. Let us consider infinitesimal transformations. The basic assumption is that
a scalar field transforms as
\begin{equation}
0=\phi'\left(x'\right)-\phi\left(x\right)=\left[\phi'\left(x'\right)-\phi\left(x'\right)\right]+\left[\phi\left(x'\right)-\phi\left(x\right)\right]\simeq\delta\phi\left(x\right)+d\phi\left(x\right),\label{scalar}
\end{equation}
where $d$ is the differential operator corresponding to $\kappa$-Poincar\'e
transformations. In order to show the invariance of the Lagrangian
appearing in (\ref{action}), it is enough to prove that
\begin{equation}
{\cal L}\left[\phi'\left(x'\right)\right]-{\cal L}\left[\phi\left(x\right)\right]\simeq\delta{\cal L}\left[\phi\left(x\right)\right]+d{\cal L}\left[\phi\left(x\right)\right]=0,\label{InvLagrangian}
\end{equation}
where $\delta{\cal L}$ is the functional variation ${\cal L}\left[\phi+\delta\phi\right]-{\cal L}\left[\phi\right]$.

The invariance of the Lagrangian is ensured if the differential satisfies
the Leibniz rule with respect to the $\star$-product,
\begin{equation}
d\left(\phi\left(x\right)\star\psi\left(x\right)\right)=\left(d\phi\left(x\right)\right)\star\psi\left(x\right)+\phi\left(x\right)\star d\psi\left(x\right),\label{leibniz}
\end{equation}
which is a standard requirement for the definition of a differential
calculus. Two different prescriptions have been proposed in the literature~\cite{Freidel:2007hk,Agostini:2006nc,AmelinoCamelia:2007uy,AmelinoCamelia:2007vj}. We adopt here the one
proposed in~\cite{Freidel:2007hk} that is based on a differential
calculus that satisfies the ``bicovariance'' property~\cite{Sitarz:1994rh}.
In this case the differential $\hat{d}$, generating infinitesimal
$\kappa$-Poincar\'e transformations in $\kappa$-Minkowski spacetime,
takes the form
\begin{equation}
\hat{d}=i\left(\hat{\epsilon}^{A}P_{A}+\hat{\omega}^{\mu\nu}L_{\mu\nu}\right)\triangleright,\label{differential}
\end{equation}
where $P_{A}$ and $L_{\mu\nu}$ are respectively the $\kappa$-Poincar\'e
translation and Lorentz generators (in classical basis). These are
defined through their action on noncommutative plane waves as $P_{A}\equiv-i\hat{\partial}_{A}$
and $L_{\mu\nu}\equiv-\frac{i}{2}\hat{x}_{[\mu}\hat{\partial}_{\nu]}\frac{\kappa}{P_{0}+P_{4}}$.
It can be proved however (see~\cite{Freidel:2007hk}) that the action
of the Lorentz generator on the field, through the Weyl map~(\ref{0.8}), reduces to the standard action
\begin{equation}
L_{\mu\nu}\triangleright\phi\left(x\right)={\cal W}^{-1}\left(L_{\mu\nu}\triangleright\phi\left(\hat{x}\right)\right)=-\frac{1}{2}x_{[\mu}\star\partial_{\nu]}\frac{\kappa}{\partial_{0}+\partial_{4}}\phi\left(x\right)=-\frac{i}{2}x_{[\mu}\partial_{\nu]}\phi\left(x\right).\label{LorentzAction}
\end{equation}
The parameters $\hat{\epsilon}^{A}$ and $\hat{\omega}^{\mu\nu}$
must obey commutation relations with $\hat{x}^{\mu}$ so that $\hat{d}$
satisfies the Leibniz rule in Minkowski spacetime
\begin{equation}
\hat{d}\left(\phi\left(\hat{x}\right)\psi\left(\hat{x}\right)\right)=\hat{d}\phi\left(\hat{x}\right)\psi\left(\hat{x}\right)+\phi\left(\hat{x}\right)\hat{d}\psi\left(\hat{x}\right).\label{leibnizNonComm}
\end{equation}
The commutation properties of $\hat{\epsilon}^{A}$ and $\hat{\omega}^{\mu\nu}$
are reported in appendix~\ref{sec:noncommParam}, and the corresponding
relations (\ref{transParStar}) and (\ref{lorParStar}) between the
images of the parameters under Weyl map and the associate $\star$-product,
lead to
\begin{equation}
\phi\left(x\right)\star\epsilon^{A}=\epsilon^{B}K_{B}^{A}\left(\partial\right)\star\phi\left(x\right),\label{transParStarPhi}
\end{equation}
and
\begin{equation}
\phi\left(x\right)\star\omega^{\mu\nu}=\Omega_{\rho\sigma}^{\mu\nu}\left(\partial\right)\omega^{\rho\sigma}\star\phi\left(x\right),\label{lorParStarPhi}
\end{equation}
where the matrices $K$ and $\Omega$ are also defined in appendix~\ref{sec:noncommParam}.

Two additional properties of the transformation parameters (see~\cite{Freidel:2007hk})
are that
\begin{equation}
\partial_{A}\epsilon^{B}=\partial_{A}\omega^{\mu\nu}=0,\label{paramIndep}
\end{equation}
and that
\begin{equation}
\left(d\phi\right)^{\dagger}=d\phi^{\dagger}.\label{complexd}
\end{equation}
We can now write the image of $\hat{d}$ under Weyl map as
\begin{equation}
d\phi\left(x\right)=\epsilon^{A}\star\partial_{A}\phi\left(x\right)+\tfrac{1}{2}\omega^{\mu\nu}\star x_{[\mu}\partial_{\nu]}\phi\left(x\right).\label{differentialStar}
\end{equation}
Using relations (\ref{transParStarPhi}) and (\ref{lorParStarPhi}),
the Lorentz action (\ref{LorentzAction}) and relations~(\ref{4.13}), (\ref{4.14}), (\ref{4.15}) and (\ref{4.16}),
one can prove that the Leibniz rule (\ref{leibniz}) is satisfied.

We can now prove the invariance of the Lagrangian for (\ref{action}).
Considering that from~(\ref{scalar}) $\delta\phi\left(x\right)=-d\phi\left(x\right)$,
the functional variation of the Lagrangian gives
\begin{equation}
\begin{split}\delta{\cal L}\left[\phi\left(x\right)\right]= & -\left(d\phi\left(x\right)\right)^{\dagger}\star\partial_{\mu}\partial^{\mu}\phi\left(x\right)-\phi^{\dagger}\left(x\right)\star\partial_{\mu}\partial^{\mu}d\phi\left(x\right)\\
 & -\left(d\phi\left(x\right)\right)\star\left(\partial_{\mu}\partial^{\mu}\phi\left(x\right)\right)^{\dagger}-\phi\left(x\right)\star\left(\partial_{\mu}\partial^{\mu}d\phi\left(x\right)\right)^{\dagger}\\
 & +m^{2}\left[\left(d\phi\left(x\right)\right)^{\dagger}\star\phi\left(x\right)+\phi^{\dagger}\left(x\right)\star d\phi\left(x\right)\right]\\
 & +m^{2}\left[\left(d\phi\left(x\right)\right)\star\phi^{\dagger}\left(x\right)+\phi\left(x\right)\star\left(d\phi\left(x\right)\right)^{\dagger}\right].
\end{split}
\label{deltaL}
\end{equation}
Given that the action (\ref{LorentzAction}) of $L_{\mu\nu}$ is the
same as the standard one, and (\ref{paramIndep}), it is straightforward
to prove that $\left[\partial^{\mu}\partial_{\mu},d\right]=0$. Indeed
the only part of $d$ on which the derivatives act is the standard
Lorentz term $\propto x_{[\rho}\partial_{\sigma]}$, so that the derivation
is the same as in the standard case:
\begin{equation}
\begin{split}\partial_{\mu}\partial^{\mu}d\phi\left(x\right)= & \epsilon^{A}\star\partial_{A}\partial_{\mu}\partial^{\mu}\phi\left(x\right)+\tfrac{1}{2}\omega^{\rho\sigma}\star\partial_{\mu}\partial^{\mu}x_{[\rho}\partial_{\sigma]}\phi\left(x\right)\\
= & \epsilon^{A}\star\partial_{A}\partial_{\mu}\partial^{\mu}\phi\left(x\right)+\tfrac{1}{2}\omega^{\rho\sigma}\star x_{[\rho}\partial_{\sigma]}\partial_{\mu}\partial^{\mu}\phi\left(x\right)+\omega^{\rho\sigma}\star\partial_{[\rho}\partial_{\sigma]}\phi\left(x\right)\\
= & d\partial_{\mu}\partial^{\mu}\phi\left(x\right).
\end{split}
\label{commBoxd}
\end{equation}
Then, using the properties (\ref{complexd}) and (\ref{leibniz}),
it follows immediately that $\delta{\cal L}\left[\phi\left(x\right)\right]=-d{\cal L}\left[\phi\left(x\right)\right]$.
We show as an example the derivation for the second row of (\ref{deltaL}).
Using the result (\ref{commBoxd}) we rewrite it first as
\[
-\left(d\phi\left(x\right)\right)\star\left(\partial_{\mu}\partial^{\mu}\phi\left(x\right)\right)^{\dagger}-\phi\left(x\right)\star\left(d\partial_{\mu}\partial^{\mu}\phi\left(x\right)\right)^{\dagger}.
\]
We now use properties (\ref{complexd}) and (\ref{leibniz}), to rewrite
it as
\[
\begin{split} & -\left(d\phi\left(x\right)\right)\star\left(\partial_{\mu}\partial^{\mu}\phi\left(x\right)\right)^{\dagger}-\phi\left(x\right)\star d\left(\partial_{\mu}\partial^{\mu}\phi\left(x\right)\right)^{\dagger}\\
= & -d\left[\phi\left(x\right)\star\left(\partial_{\mu}\partial^{\mu}\phi\left(x\right)\right)^{\dagger}\right].
\end{split}
\]

We have thus shown under which hypotheses the action is $\kappa$-Poincar\'e
invariant. To conclude, let us discuss briefly what the condition
(\ref{scalar}) implies for the transformation of the field. If we
consider the Fourier transform of the field generically as
\begin{equation}
\phi\left(x\right)=\int d\mu\left(p\right)\tilde{\phi}\left(p\right)e^{-ip\cdot x},
\end{equation}
then
\[
\begin{split}d\phi\left(x\right)= & \epsilon^{A}\star\int d\mu\left(p\right)\tilde{\phi}\left(p\right)\partial_{A}e^{-ip\cdot x}+\tfrac{1}{2}\omega^{\mu\nu}\star\int d\mu\left(p\right)\tilde{\phi}\left(p\right)x_{[\mu}\partial_{\nu]}e^{-ip\cdot x}\\
= & -i\epsilon^{A}\star\int d\mu\left(p\right)\tilde{\phi}\left(p\right)p_{A}e^{-ip\cdot x}-\tfrac{1}{2}\omega^{\mu\nu}\star\int d\mu\left(p\right)\tilde{\phi}\left(p\right)p_{[\mu}\frac{\partial}{\partial p^{\nu]}}e^{-ip\cdot x}\\
= & \int d\mu\left(p\right)\left(-ip_{A}\tilde{\phi}\left(p\right)\epsilon^{A}+\tfrac{1}{2}p_{[\mu}\frac{\partial}{\partial p^{\nu]}}\tilde{\phi}\left(p\right)\omega^{\mu\nu}\right)\star e^{-ip\cdot x}.
\end{split}
\]
The field variation $\delta\phi=-d\phi$ implies that formally we
can state
\begin{equation}
\delta\tilde{\phi}\left(p\right)=\left(i\epsilon^{A}p_{A}\tilde{\phi}\left(p\right)-\tfrac{1}{2}\omega^{\mu\nu}p_{[\mu}\frac{\partial}{\partial p^{\nu]}}\tilde{\phi}\left(p\right)\right)\star.
\end{equation}
The last relation is very similar to its classical analogous, which
is given by\footnote{We are here using notations such that for a finite Poincar\'e transformation
$\phi'\left(x'\right)=U^{-1}\left(\Lambda,a\right)\phi\left(\Lambda x+a\right)U\left(\Lambda,a\right)\simeq\phi\left(x\right)+\delta\phi\left(x\right)+d\phi\left(x\right)$,
so that $U^{-1}\left(\Lambda,a\right)\phi\left(x\right)U\left(\Lambda,a\right)\simeq\delta\phi\left(x\right)=-d\phi\left(x\right)$.}
\begin{equation}
U^{-1}\left(\Lambda,a\right)\tilde{\phi}\left(p\right)U\left(\Lambda,a\right)=e^{i\left(\Lambda^{-1}a\right)\cdot p}\tilde{\phi}\left(\Lambda^{-1}p\right)\simeq\tilde{\phi}\left(p\right)+i\epsilon^{\mu}p_{\mu}\tilde{\phi}\left(p\right)-\tfrac{1}{2}\omega^{\mu\nu}p_{[\mu}\frac{\partial}{\partial p^{\nu]}}\tilde{\phi}\left(p\right).
\end{equation}

\section{Properties of the noncommutative parameters $\epsilon^{A}$ and $\omega^{\mu\nu}$}
\label{sec:noncommParam}

The properties of the noncommutative parameters are derived in \cite{Freidel:2007hk}.
For the translation parameter they amount to
\[
\left[\hat{x}^{\mu},\hat{\epsilon}^{A}\right]=\left(X^{\mu}\right)_{\ B}^{A}\hat{\epsilon}^{B},
\]
where
\[
\hat{X}^{0}=-\frac{i}{\kappa}\left(\begin{array}{ccc}
0 & {\bf 0}^{T} & 1\\
{\bf 0} & {\bf 0}_{3\times3} & {\bf 0}\\
1 & {\bf 0}^{T} & 0
\end{array}\right),\qquad\hat{{\bf X}}=\frac{i}{\kappa}\left(\begin{array}{ccc}
0 & {\bf n}^{T} & 0\\
{\bf n} & {\bf 0}_{3\times3} & {\bf n}\\
0 & -{\bf n}^{T} & 0
\end{array}\right),
\]
where ${\bf n}$ is a unit vector in standard basis. In terms of plane
waves, they satisfy the relation
\begin{equation}
\hat{e}_{k}\hat{\epsilon}^{A}\hat{e}_{k}^{-1}=\hat{\epsilon}^{B}K_{B}^{A}\left(p\left(k\right)\right),\label{transParNC}
\end{equation}
with
\[
K\left(p\right)=\frac{1}{\kappa}\left(\begin{array}{ccc}
p_{4}+\frac{{\bf p}^{2}}{p_{0}+p_{4}} & -\frac{\kappa}{p_{0}+p_{4}}{\bf p}^{T} & p_{0}\\
-{\bf p} & \kappa{\bf 1}_{3\times3} & -{\bf p}\\
p_{0}-\frac{{\bf p}^{2}}{p_{0}+p_{4}} & \frac{\kappa}{p_{0}+p_{4}}{\bf p}^{T} & p_{4}
\end{array}\right),
\]
\[
K^{-1}\left(p\right)=\frac{1}{\kappa}\left(\begin{array}{ccc}
p_{4}+\frac{{\bf p}^{2}}{p_{0}+p_{4}} & {\bf p}^{T} & -p_{0}+\frac{{\bf p}^{2}}{p_{0}+p_{4}}\\
\frac{\kappa}{p_{0}+p_{4}}{\bf p} & \kappa{\bf 1}_{3\times3} & \frac{\kappa}{p_{0}+p_{4}}{\bf p}\\
-p_{0} & -{\bf p}^{T} & p_{4}
\end{array}\right).
\]
Notice that $\hat{X}^{\mu}$ and $K\left(p\left(k\right)\right)$
matrices coincide respectively with the 5D representations of $\hat{x}^{\mu}$
and $\hat{e}_{k}$ given in \textbf{(2)} and \textbf{(4)}, in agreement
with the fact that $\hat{\epsilon}^{A}$ form a representation of
$\kappa$-Minkowski algebra. For the Lorentz parameter the commutation
properties are given by
\begin{equation}
\hat{e}_{k}\hat{\omega}^{\mu\nu}\hat{e}_{k}^{-1}=\hat{\omega}^{\rho\sigma}\Omega_{\rho\sigma}^{\mu\nu}\left(p\left(k\right)\right),\label{lorParNC}
\end{equation}
with
\[
\Omega_{\rho\sigma}^{\mu\nu}\left(p\right)=\delta_{[\rho}^{\,\mu}\tau_{\sigma]}^{\,\nu}\left(p\right)
\]
and
\[
\tau\left(p\right)=\left(\begin{array}{cc}
2\frac{\kappa}{p_{0}+p_{4}}-1\quad & -2\frac{{\bf p}}{p_{0}+p_{4}}\\
0 & {\bf 1}
\end{array}\right).
\]

Using the (inverse) Weyl map \eqref{0.12} with eqs. (\ref{transParNC})
and (\ref{lorParNC}), we obtain the corresponding properties for
the $\star$-product between the parameters and the plane waves
\begin{equation}
\begin{gathered}e_{p}\star\epsilon^{A}=K_{B}^{A}\left(p\right)\epsilon^{B}\star e_{p},\\
\epsilon^{A}\star e_{p}=\left(K^{-1}\right)_{B}^{A}\left(p\right)e_{p}\star\epsilon^{B},
\end{gathered}
\label{transParStar}
\end{equation}
and
\begin{equation}
\begin{gathered}e_{p}\star\omega^{\mu\nu}=\Omega_{\rho\sigma}^{\mu\nu}\left(p\right)\omega^{\rho\sigma}\star e_{p},\\
\omega^{\mu\nu}\star e_{p}=\left(\Omega^{-1}\right)_{\rho\sigma}^{\mu\nu}\left(p\right)e_{p}\star\omega^{\rho\sigma}.
\end{gathered}
\label{lorParStar}
\end{equation}

\end{document}